\documentclass[%
 reprint,
nofootinbib,
 amsmath,amssymb,
 aps,
onecolumn
]{revtex4-1}

\usepackage{graphicx}
\usepackage{dcolumn}
\usepackage{bm}

\usepackage{amsmath}
\usepackage{amsfonts}
\usepackage{amssymb}
\usepackage{multirow}
\usepackage{array}
\usepackage{subfigure}
\usepackage{sidecap}

\begin{document}

\title{Same-Sign Dileptons from Colored Scalars\\
 in the Flavorful Top-Coloron Model}

\author{R. Sekhar Chivukula, Elizabeth H. Simmons, and Natascia Vignaroli}
 \affiliation{Department of Physics and Astronomy, Michigan State  University, East Lansing, MI 48824, USA.}

\date{\today}

\begin{abstract}
In this paper we study the phenomenology of color-octet and color-singlet scalars in the flavorful Top-Coloron model.
We discuss  the relevant production mechanisms at hadron colliders and the dominant decay modes, highlighting the most promising signatures for discovery, and derive bounds on the masses of the new scalars from 
LHC and Tevatron data. Of particular interest is the case in which color-octet scalars are pair produced
and each decay to $t\bar{c}$ or $\bar{t}c$, leading to a same-sign-dilepton final state. LHC data places a lower limit of 440 GeV on the octet mass in this scenario.  The case of an octet lighter than the top, where the octet only decays into jets, has been tested by the Tevatron, which excludes the mass region from 50 to 125 GeV. The 8 TeV LHC is not yet sensitive to the observation of  the color-singlet states, which are produced at rates much smaller than the octets. Nevertheless, the color-singlet pseudoscalar can be discovered at the 14 TeV LHC by analyzing the channel where it is produced from the decay of a color-octet vector boson.
\end{abstract}

\maketitle


\section{Introduction}

New colored scalars are a general prediction in a wide class of models with an extended color sector \cite{Hill:2002ap, Hill:1991at, Chivukula:1996yr, Frampton:1987dn}, grand-unified theories \cite{Perez:2008ry,FileviezPerez:2008ib,Bertolini:2013vta} and even certain extra-dimensional models \cite{Burdman:2006gy}.
Their phenomenology is of particular interest for hadron colliders, which can copiously produce colored particles. Colored scalars are also receiving increased attention because of their possible effect on Higgs decay rates (see, for example, \cite{Kumar:2012ww,Kribs:2012kz,Cao:2013wqa,Chang:2012ta,Dobrescu:2011aa,Manohar:2006ga,Arsham}). 

Previous studies of the collider phenomenology of colored scalars have considered generic interactions of these latter with Standard Model quarks, without referring to a specific flavor structure \cite{Dobrescu:2007yp,Bai:2010dj}, or have  focused on scenarios of minimal-flavor-violation \cite{Manohar:2006ga,Gresham:2007ri,Arnold:2011ra,Gerbush:2007fe,Carpenter:2011yj}. Here we will study the phenomenology of colored scalars in a scenario of next-to-minimal flavor violation, which leads to novel and promising signatures  for searches at colliders. Specifically, we study the flavorful Top-Coloron model, that we have recently introduced in \cite{Chivukula:2013kw}, which naturally addresses the experimental observation that the third family of quarks has only a small mixing with the lighter families. In this model the new scalars appear in the spectrum because an extended color sector, $SU(3)_1\times SU(3)_2$, is broken down to the QCD group when a $(3,\bar{3})$ scalar acquires a vacuum expectation value.  Below the symmetry-breaking scale, the new physical scalar states are a color-octet scalar, a color-singlet pseudoscalar and a color-singlet scalar.  Due to the model's flavor structure, limits on the scalars from flavor observables are quite weak, as described in Appendix \ref{Sec:flavor}.  In fact, the scalars can be relatively light and thus their phenomenology at the LHC and the Tevatron is very interesting.

We show that the color-octet scalar is dominantly produced at hadron colliders in pairs. If it is lighter than the top quark, it decays into jets, a case that is constrained by a recent CDF search \cite{Aaltonen:2013hya}, which excludes the mass region from 50 to 125 GeV. If the octet is heavier than the top, it decays dominantly into a top plus a charm, with the same decay rate into $t\bar{c}$ or $\bar{t}c$ channels. In this case, pair-production can lead to a same-sign-dilepton final state, a very promising signature for discovery. We find that the searches for new physics at the 8 TeV LHC in this channel already place a lower limit of 440 GeV on the octet mass. 

The color-singlet states cannot be produced at tree level through QCD interactions. They can be produced singly at one-loop order, but we find that the corresponding rates are too small to be observable at the LHC. The current searches at colliders therefore do not place direct bounds on the masses of the color-singlet states.  The color-singlet  pseudoscalar can, however, be produced at tree-level from the decay of a color-octet vector boson, which we generically denote as a  ``coloron''. This proves to be a promising search channel for the LHC at a center-of-mass energy of 14 TeV.

The paper is organized as follows. We review the flavorful Top-Coloron model and introduce the scalar potential and spectrum in Sec. \ref{Sec:review}. The phenomenology of the color-octet scalar is discussed in Sec. \ref{sec:pheno-octet}; for the physically distinct octet mass ranges, we show the relevant production mechanisms and the dominant decay modes and we highlight the most promising signatures for searches at colliders. We also derive the bounds on the octet mass coming from current searches at the Tevatron and at the LHC. Sec. \ref{Sec:singlet} describes the phenomenology of the color-singlet pseudoscalar. Promising channels for the singlet pseudoscalar searches, where the singlet is produced from the decay of a coloron are discussed in Sec. \ref{Sec:coloron}. Sec. \ref{Sec:phiR} describes the phenomenology of the color-singlet scalar. We will draw our conclusions and comments in Sec. \ref{Sec:conclusion}.

\section{The Flavorful Top-Coloron Model }\label{Sec:review}

\subsection{Gauge and Fermion Sectors}

In \cite{Chivukula:2013kw}, we introduced a renormalizable model of an extended color gauge sector which naturally address the experimental observation that the third family of quarks has only a small mixing with the lighter families, and which  realizes next-to-minimal flavor violation. The color interactions are extended to an $SU(3)_1 \times SU(3)_2$ structure in which the third generation quarks couple differently than the lighter quarks. In addition to a set of heavy color-octet vector bosons (colorons), the model also contains a set of heavy weak vector quarks. Mixing between ordinary quark generations occurs only because all three generations mix with the vector quarks. Since the heavy vector quarks
transform in the same way under the extended color sector as the third generation quarks and differently from the lighter families, we naturally obtain small mixing between the third generation of quarks and the first two.

The model has the gauge structure 
$SU(3)_1 \times SU(3)_2 \times SU(2)_W \times U(1)_Y$. The  $SU(3)_1 \times SU(3)_2$ gauge couplings, $g_1$ and $g_2$, are related to the QCD coupling $g_s$ through
\begin{equation}
g_s=g_1\sin\omega=g_2\cos\omega ,
\end{equation}
where $\omega$ is a new gauge mixing angle. Gauge symmetry breaking
occurs in two steps:
\begin{itemize}
\item $SU(3)_1 \times SU(3)_2 \to SU(3)_C$ due to the (diagonal) expectation value
$\langle \Phi \rangle \propto u \cdot {\cal I}$, where the scalar, $\Phi$, transforms as a $(3,\bar{3})$ under
$SU(3)_1 \times SU(3)_2$ and ${\cal I}$ is the identity matrix,

\item $SU(2)_W \times U(1)_Y \to U(1)_{em}$ in the usual way due to a Higgs doublet
$\phi$ transforming as a $2_{1/2}$ of the electroweak group, and with the usual
vacuum expectation value given by $v \approx 246$ GeV.

\end{itemize}

We assume that coloron symmetry breaking occurs at a scale large compared
to the weak scale, $u \gg v$. The mass-squared matrix for the $SU(3)_1 \times SU(3)_2$ gauge bosons $A^{a}_{1\mu}$ and $A^{a}_{2\mu}$ is given by

\begin{equation}
-\frac{1}{2} u^2 
\begin{pmatrix}
g^2_1 & -g_1 g_2 \\
-g_1 g_2 & g^2_2
\end{pmatrix}
\end{equation}
Diagonalizing this matrix reveals the gluon and the coloron mass eigenstates $G^a$ and $C^a$
\begin{align}
\begin{split}
G^a_{\mu} = \sin\omega A^a_{1\mu} + \cos\omega A^a_{2\mu}~, \\
C^a_{\mu} = \cos\omega A^a_{1\mu} - \sin\omega A^a_{2\mu}~,
\end{split}
\end{align}
with masses
\begin{equation}
M_G=0~, \qquad M_C=\frac{g_S \, u}{\sin\omega \cos\omega} \ .
\end{equation}
The gluon and coloron, respectively, couple to the following currents:
\begin{align}
\begin{split}
& g_S J^{a \mu}_G = g_S ( J^{a \mu}_1 +  J^{a \mu}_2)~, \\
& g_S J^{a \mu}_C = g_S ( \cot\omega J^{a \mu}_1 - \tan\omega  J^{a \mu}_2)~, 
\end{split}
\end{align}
where $J^{a \mu}_i $ denotes the color current associated with $SU(3)_i$.

\begin{table}
\centering
\begin{tabular}{|c|c|c|c|c|}
\hline
Particle & $SU(3)_1$ & $SU(3)_2$ & $SU(2)$ & $U(1)$ \\
\hline\hline
$\vec{\mathcal Q}_L= \begin{pmatrix}
 q_L \\
 Q_L
 \end{pmatrix}$ & 3 & 1 & 2 & +1/6\\
\hline
$t_R$ & 3 & 1 & 1 & +2/3\\
\hline
$b_R$ & 3 & 1 & 1 & -1/3\\
\hline
$Q_R$ & 3 & 1 & 2 & +1/6\\
\hline\hline
$\vec{\psi}_L = \begin{pmatrix}
 \psi^1_L \\
 \psi^2_L
 \end{pmatrix}
$ & 1 & 3 & 2 & +1/6\\
\hline
$\vec{u}_R= \begin{pmatrix}
 u^1_R \\
 u^2_R
 \end{pmatrix}$ & 1 & 3 & 1 & +2/3\\
\hline
$\vec{d}_R= \begin{pmatrix}
 d^1_R \\
 d^2_R
 \end{pmatrix}$ & 1 & 3 & 1 & -1/3\\
\hline\hline
$\phi$ & 1 & 1 & 2 & +1/2\\
\hline
$\Phi$ & 3 & $\bar{3}$ & 1 & 0 \\
\hline\hline
\end{tabular}
\caption{$SU(3)_1 \times SU(3)_2 \times SU(2) \times U(1)$ gauge charges of
the particles in this model. The $\phi$ and $\Phi$, respectively, denote the scalars responsible
for breaking the electroweak and (extended) strong sectors, while all other listed 
particles are fermions. The vectors ($\vec{\psi}_L$, $\vec{u}_R$,
$\vec{d}_R$, and $\vec{\mathcal Q}_L$) denote different fermion flavors with the
same gauge charges, where the superscripts [1,2] refer to the two light fermion generations. \label{table:i}}
\end{table}

The matter fields of the model, which are summarized in Table \ref{table:i}, include the chiral quark generation ($q_L$, $t_R$ and $b_R$), associated primarily with the third generation quarks, and one vectorial quark generation ($Q_{L,R}$), coupled to $SU(3)_1$. The two remaining chiral quark generations ($\vec{\psi}_L$, $\vec{u}_R$, $\vec{d}_R$), associated primarily with the two light quark generations, are coupled to $SU(3)_2$. The flavor properties of the model depend on the fermion masses and Yukawa couplings.
The Yukawa couplings for the light two generations are given by\footnote{The factors of $M/v$ in the Yukawa couplings are introduced for convenience in the analysis of quark mixings.}
\begin{equation}
\frac{\sqrt{2}M}{v} \cdot \left(
\vec{\bar{\psi}}_L \phi \lambda_d \vec{d}_R + \vec{\bar{\psi}}_L \tilde{\phi} \lambda_u \vec{u}_R
\right) +\text{ H.c.}~,
\end{equation}
where $M$ is the Dirac mass for vector heavy quark and $\lambda_{u,d}$ are $2\times 2$ complex matrix Yukawa couplings. The third-generation quark Yukawa couplings are given by
\begin{equation}
\frac{\sqrt{2}M}{v} \cdot \left(
\beta_b \bar{q}_{L} \phi  b_R + \beta_t \bar{q}_{L} \tilde{\phi} t_R
\right) + \text{H.c.}~,
\end{equation}
where the $\beta_{t,b}$ can be chosen to be real. Neglecting the (small) mixing of the third-generation quarks with the first two generations, the parameters $\lambda_{u,d}$ and $\beta_{t,b}$ are just equal to the corresponding parameters in the standard model, up to the factor of $\sqrt{2}M/v$.
Mixing between the third-generation quarks and the vector quarks occurs
through
\begin{equation}
\frac{\sqrt{2}M}{v} \cdot \left(  \lambda'_b \bar{Q}_{L}\phi b_R+\lambda'_t \bar{Q}_{L}\tilde{\phi} t_R\right) + \text{H.c.}~,
\end{equation}
where $\lambda^{'}_{b,t}$ are complex numbers, and mixing between the first- and second-generation quarks and the vector quarks occurs through the Yukawa
couplings to the $\Phi$ field:
\begin{equation}\label{eq:Phi-term}
\frac{M}{u}\cdot \left(
\vec{\bar{\psi}}_L \cdot \vec{\alpha}\, \Phi Q_R
\right) + \text{H.c.}~.
\end{equation}
where $\vec{\alpha}$ is a two component complex vector, 
\begin{eqnarray}
\vec{\alpha}\equiv \left( \begin{array}{c}   \alpha_1 \\   \alpha_2 \end{array} \right)\ .
\end{eqnarray}

The full flavor analysis of this model is presented in \cite{Chivukula:2013kw}.  Here, we summarize the results that are most relevant to  the phenomenology of the color-octet and color-singlet scalars. First, we find that the CKM matrix is correctly reproduced by:
\begin{align}
\begin{split}\label{est-ckm}
& V_{ub}=\alpha_1 \left(\frac{\lambda'_t}{\beta_t}-\frac{\lambda'_b}{\beta_b}\right)=A\lambda^3(\rho-i\eta)=0.00131-i0.00334\\
& V_{cb} = \alpha_2 \left(\frac{\lambda'_t}{\beta_t}-\frac{\lambda'_b}{\beta_b}\right)=A\lambda^2=0.0415
\end{split}
\end{align} 
where \cite{Beringer:1900zz}:
\begin{align}
\lambda=0.22535 &\pm 0.00065 \\
A=0.817 & \pm 0.015\\
\rho = 0.140 & \pm 0.018 \\
\eta = 0.357 & \pm 0.014
\end{align}
are the CKM parameters in the Wolfenstein parameterization. The model thus consistently reproduces the observed pattern of quark mixing if $\alpha_2$ is $\mathcal{O}(\lambda^2)$, $\alpha_1$ is $\mathcal{O}(\lambda^3)$ and both $\lambda^{'}_{t}/\beta_t$ and $\lambda^{'}_{b}/\beta_b$ are order 1.

Even though the model is consistent with CKM mixings, additional contributions to various flavor-changing neutral currents (FCNC) arise from non-standard interactions, namely from the couplings of the coloron to fermions and from the mixing of the ordinary quarks with the new weak vector fermions.\footnote{There are also additional contributions from scalar exchange. As we show in Appendix \ref{Sec:flavor}, however, these contributions are suppressed by the Yukawa couplings and are therefore very small.} These additional contributions are controlled in the down quark sector by the matrices
\begin{equation}\label{matrix:DL4}
{\mathcal D}_L = \begin{pmatrix} V_{ud} & V_{us} & 0 & \alpha_1  \\  V_{cd} & V_{cs} & 0 & \alpha_2 \\
0 & 0 & 1 & 0 \\-(V_{ud}\alpha^*_1+V_{cd}\alpha^*_2) & -(V_{us}\alpha^*_1+V_{cs}\alpha^*_2) & 0 &1 \end{pmatrix}~,
\end{equation}
\begin{equation}\label{matrix:DR4}
{\mathcal D}_R=\begin{pmatrix} 1 & 0 & 0 & 0  \\  0 & 1 & 0 & 0  \\
0 & 0 & 1 & \lambda^{'*}_b \\
0 & 0 & -\lambda^{'}_b &1 \end{pmatrix}~,
\end{equation}  
in a basis where the first two components are the light-quark fields, the next is the third generation, and the last represents the vectorlike quarks. Similarly, in the up quark sector the additional contributions are
controlled by the matrices ${\mathcal U}_{L,R}$, which are obtained from the ${\mathcal D}_{L,R}$ by setting $\lambda^{'}_b \to \lambda^{'}_t$, and $V_{ud}, V_{cs}\to 1$ and  $V_{us}, V_{cd}\to 0$.

The strongest constraints on the model parameters are placed by data on $b\to s \gamma$ and $\Delta F=2$ meson mixing processes, especially $K$ meson mixing. Data on $b\to s \gamma$ leads to the bound
\begin{equation}\label{eq:bsgamma-lim}
\left|\frac{\alpha_2 \lambda^{'}_b}{\beta_b}\right|<0.0085 \ .
\end{equation}
The constraint from an additional contribution to CP-violation in $K$ meson mixing gives a bound on the coloron mass:
\begin{equation}
M_C > 3.6 \, g_S \,  \left(\frac{|\alpha_2|}{\lambda^2}\right)^2 \cot\omega \; \text{TeV} \ .
\end{equation}

Further limits on model parameters come from direct searches at colliders for colorons and heavy vector quarks. 
The branching ratios for the decays of heavy vector quarks, in the limit $\lambda^{'}_b\ll \lambda^{'}_t$, read:
\begin{align}
\begin{split}\label{eq:BR-Q}
& BR(Q^d \to W t_R) \simeq 1 \\
& BR(Q^u \to Z t_R) \simeq BR(Q^u \to h t_R) \simeq 0.5 \\
\end{split} 
\end{align}
The strongest bound on the vector quark mass comes from the ATLAS search \cite{atlas-conf-2012-130} for pair produced 4th generation down-type quarks, decaying predominantly to $Wt$, which places a limit on the $Q^d$ mass:
\begin{equation}\label{eq:limit-Qd}
M_{Q^d}\gtrsim 670 \, \text{GeV} \ ,
\end{equation}
where 
\[
M_{Q^{d,u}}=M+\frac{1}{2}\left(|\lambda^{'}_{b,t}|^2+|\vec{\alpha}|^2 \right) M \ .
\]
Note that $M_{Q^d}\simeq M_{Q^u}\simeq M $. We will discuss coloron phenomenology, in the case where the coloron is heavier than color-octet and color-singlet scalars, in Sec. \ref{Sec:coloron}.

\subsection{Scalar Sector}
\label{sec:potential}

The new color-octet and color-singlet scalars emerge from the symmetry breaking $SU(3)_1 \times SU(3)_2 \to SU(3)_C$ induced by the expectation value of the scalar $\Phi$. Their properties can be understood by analyzing the $\Phi$ potential, as we will do in this section adapting the notation and analysis given in \cite{Bai:2010dj}.

The most general renormalizable potential for $\Phi$ is:
\begin{equation}
\label{eq:potential}
V(\Phi)=-m^2_{\Phi}\text{Tr}(\Phi\Phi^\dagger) -\mu (\text{det}\Phi+\text{H.c.})+\frac{\xi}{2}\left[ \text{Tr}(\Phi\Phi^\dagger) \right]^2+\frac{k}{2}\text{Tr}(\Phi\Phi^\dagger\Phi\Phi^\dagger) \ ,
\end{equation}
where 
\begin{equation}
\text{det} \Phi = \frac{1}{6}\epsilon^{ijk}\epsilon^{i'j'k'}\Phi_{ii'}\Phi_{jj'}\Phi_{kk'} \ ,
\end{equation}
and where, without loss of generality, we choose $\mu >0$.
We assume $m^2_\Phi >0$ so that $\Phi$ acquires a (positive) diagonal expectation value:
\begin{equation}
\langle \Phi \rangle = u \cdot \mathcal{I} =\frac{\sqrt{4(k+3\xi)m^2_{\Phi}+\mu^2}+\mu}{2(k+3\xi)}\cdot \mathcal{I} \ .
\end{equation}
This implies:
\begin{equation}
\label{eq:u-lim}
u\geq \frac{\mu}{k+3\xi}\ .
\end{equation}

By expanding $\Phi$ around the vacuum, we obtain:
\begin{equation}\label{eq:Phi}
\Phi=u+\frac{1}{\sqrt{6}}\left(\phi_R+i\phi_I\right)+\left(G^a_H+iG^a_G\right)T^a \ ,
\end{equation}
where $\phi_R$ and $\phi_I$ are singlets under $SU(3)_C$, while $G^a_H$ and $G^a_G$, $a=1,\dots,8$, are color octets. In particular, $G^a_G$ are the Nambu-Goldstone bosons associated with the breaking $SU(3)_1 \times SU(3)_2 \to SU(3)_C$ and become the longitudinal polarization modes of the colorons.  The spectrum of the physical scalar fields is largely determined by the symmetries of the scalar potential.
In the limit $\mu \to 0$ the potential has a global symmetry $U(1)_1 \times U(1)_2$. The singlet pseudoscalar $\phi_I$ is the Nambu-Goldstone boson associated with the breaking by $\langle \Phi \rangle$ of $U(1)_1 \times U(1)_2$ down to the diagonal $U(1)$ subgroup. The squared mass of the $\phi_I$ is given by
\begin{equation}\label{eq:MphiI}
M^2_{\phi_I}=3\mu u \ .
\end{equation}
In the limit $k \to 0$ and $\mu \to 0$, there is a global $SO(18)$ symmetry, which is spontaneously broken by $\langle \Phi \rangle$ down to $SO(17)$. The states $G_H$ and $\phi_I$ are the associated extra Nambu-Goldstone bosons. The squared mass of $G_H$ is thus proportional to $\mu$ and $ku$ and we find
\begin{equation}\label{eq:MGH}
M^2_{G_H}=2\mu u + 2 k u^2=\frac{2}{3}M^2_{\phi_I}+2ku^2\ .
\end{equation}
The squared mass of $\phi_R$ is given by
\begin{equation}\label{eq:MphiR}
M^2_{\phi_R}=2 (k+3\xi)u^2-\frac{M^2_{\phi_I}}{3}\ 
\end{equation}
and the condition in Eq. (\ref{eq:u-lim}) guarantees $M^2_{\phi_R}\geq 0$. \\

The potential is stable provided that
\begin{equation}\label{eq:stability}
k+3\xi>0 \ ,
\end{equation}
and hence, from Eq. (\ref{eq:u-lim}),
\begin{equation}
\label{eq:u-lim2}
u\geq \mu \ .
\end{equation}
Note that $k$ can be negative -- but, from Eq. (\ref{eq:MGH}), we have
\begin{equation}
k > -\,\frac{\mu}{u}>-1~.
\end{equation}

For $\mu$ and $u$ of the same order of magnitude, and $k$ and $\xi$ of the same order of magnitude, we 
expect 
\begin{equation}
M^2_{G_H}\simeq M^2_{\phi_I} \simeq M^2_{\phi_R} \ .
\end{equation}
A departure from this natural spectrum implies a tuning of the $ku$, $\xi u$ and $\mu$ parameters or a limit where one of the parameters is much greater (or much smaller) than the others. For example, in the limit $\mu \ll u$, one finds
\begin{equation}
M^2_{\phi_I} \ll M^2_{G_H}~, M^2_{\phi_R}~.
\end{equation}
Alternatively, if we set $ku\simeq -\mu$, then
\begin{equation}
M^2_{\phi_I} \gg M^2_{G_H}~, M^2_{\phi_R}\ .
\end{equation}
In the following analysis we will consider the most general spectrum for the scalars. Among other things, we will see that the scenario $M^2_{\phi_I} \gg M^2_{G_H},\,M^2_{\phi_R}$ is more difficult to discover at hadron colliders. Nevertheless, we will show that a portion of the parameter space for this limiting case has already been excluded by ATLAS and CMS data.

\section{octet phenomenology}
\label{sec:pheno-octet}

In this section we study the phenomenology of the color-octet scalar $G_H$. We first analyze its production mechanisms and decays. Then we show, depending on the octet mass, the dominant decay modes and the most promising signatures for searches at colliders and derive the exclusion regions based on existing LHC and Tevatron data.

\subsection{Production}

\begin{figure}
\includegraphics[width=.9\textwidth]{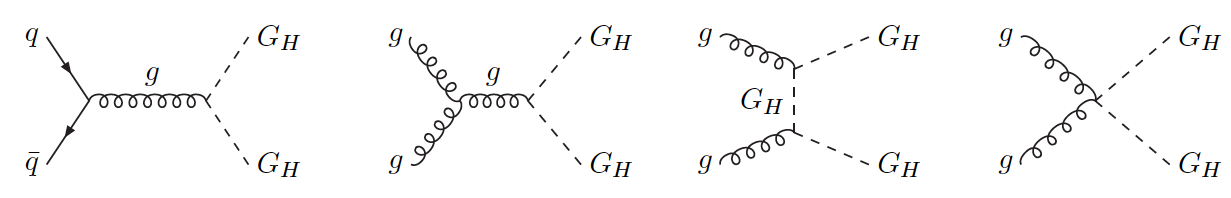}
\caption{Feynman diagram for pair production of color-octet scalars, $G_H$. }
\label{fig:dia-GH-double}
\end{figure}

\begin{figure}
\includegraphics[width=.6\textwidth]{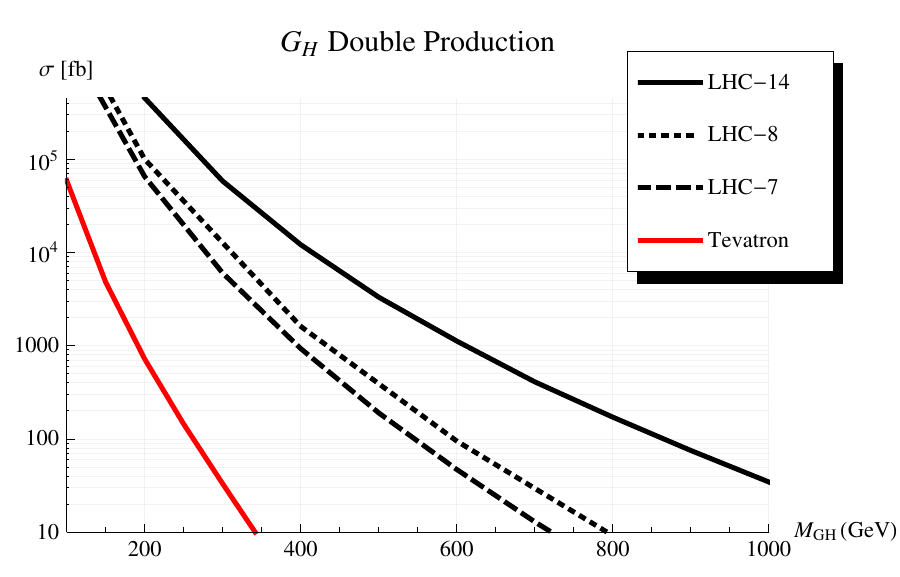}
\caption{Cross sections for scalar octet pair-production at the Tevatron (lowest curve) and at the 7, 8 and 14 TeV LHC (upper curves, as in the key). }
\label{fig:xsec}
\end{figure}

The color-octet scalars, $G_H$, are mainly produced in pairs through their interactions with gluons (Fig. \ref{fig:dia-GH-double}):
\begin{equation}\label{eq:GH-gluons}
\frac{g^2_s}{2}f^{abc}f^{ade}G^b_{\mu}G^{\mu d}G^c_H G^e_H +g_s f^{abc} G^a_{\mu} G^b_H \partial^{\mu} G^c_H
\end{equation}
Fig. \ref{fig:xsec} shows the cross sections for pair production at the Tevatron and at the 7, 8 and 14 TeV LHC, obtained from \uppercase{MadGraph} \cite{Alwall:2011uj}. 

\begin{figure}
\includegraphics[width=.7\textwidth]{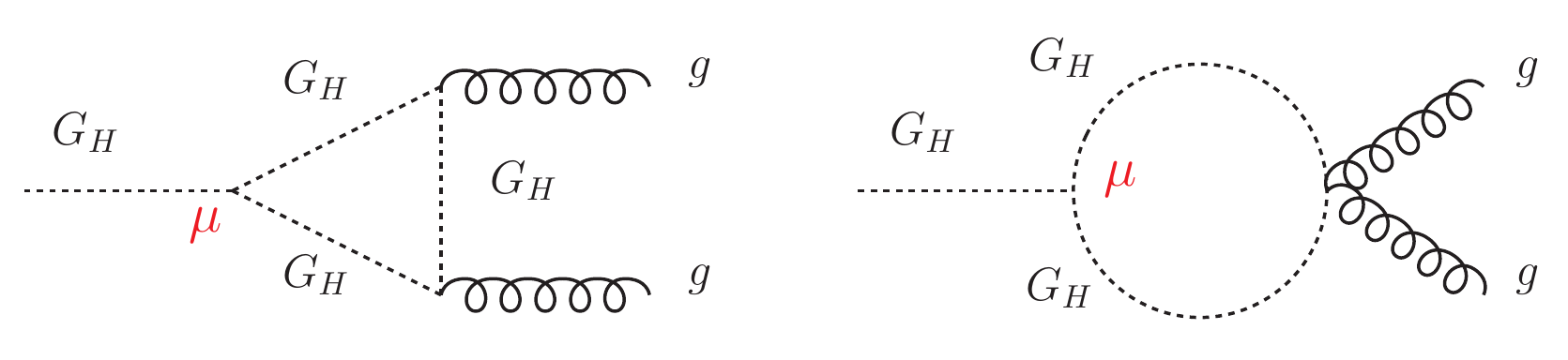}
\caption{Loop-exchange of color-octet scalars contributing to $G_H$ single-production and to $G_H$ decays to gluons.}
\label{fig:dia-Gamma0}
\end{figure}

\begin{figure}
\includegraphics[width=.4\textwidth]{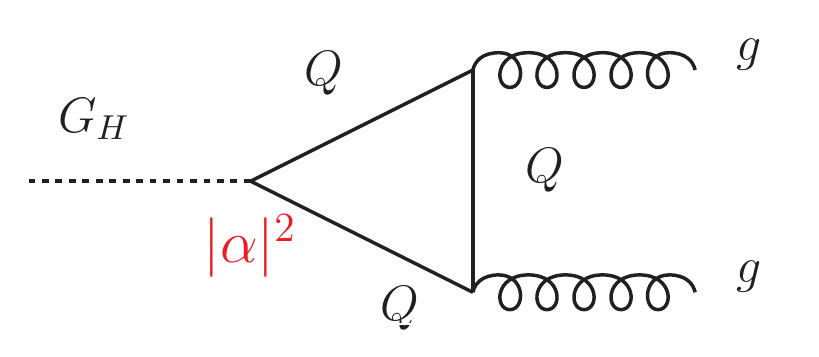}
\caption{Loop-exchange of vectorlike quarks contributing to $G_H$ single-production and to $G_H$ decays to gluons. This contribution is suppressed relative to the contributions illustrated in Fig. \ref{fig:dia-Gamma0}, as discussed in the text.}
\label{fig:dia-GammaQ}
\end{figure}

The color octets can also be produced singly via gluon-gluon fusion, as shown in Fig. \ref{fig:dia-Gamma0}. This occurs at one-loop order through the cubic interaction
\begin{equation}\label{eq:GH-cubic}
\frac{\mu}{6} d_{abc} G^a_H G^b_H G^c_H   
\end{equation}
which arises from the $\mu(\det\Phi+\text{H.c.})$ term in the potential (\ref{eq:potential}); where $d_{abc}$ is the SU(3) totally symmetric tensor. Overall, single production of $G_H$ can be described by the effective coupling \cite{Gresham:2007ri}
\begin{equation}\label{eq:single-eff}
-\frac{1}{4} C_{ggG} d_{abc} G^a_{\mu\nu} G^{\mu\nu b} G^c_H
\end{equation}
with
\begin{equation}\label{eq:cGHg}
C_{ggG}=\sqrt{\frac{1}{6}}\frac{\alpha_s}{\pi }\frac{\mu}{M^2_{G_H}}\left(\frac{\pi^2}{9}-1\right)=\sqrt{\frac{1}{6}}\frac{\alpha_s}{3\pi u}\left( \frac{M_{\phi_I}}{M_{G_H}}\right)^2\left(\frac{\pi^2}{9}-1\right)\ \ ,
\end{equation}
where we have used eq. (\ref{eq:MphiI}).\footnote{There is also a one-loop contribution to the single production of $G_H$ given by the exchange of vectorlike quarks (Fig. \ref{fig:dia-GammaQ}). As we will demonstrate in the next section, this contribution is subleading due to an $|\alpha|^4$ suppression and can thus be safely neglected.}
We calculate the rate for single production of $G_H$ by implementing the above interaction (\ref{eq:single-eff}) in \uppercase{MadGraph}. We obtain cross sections that are typically smaller than those for pair production of $G_H$ by a factor of the order of 10$^4$. The octet single-production rate is thus negligible compared to the double-production rate.

\subsection{Decay}

\begin{figure}
\includegraphics[width=.8\textwidth]{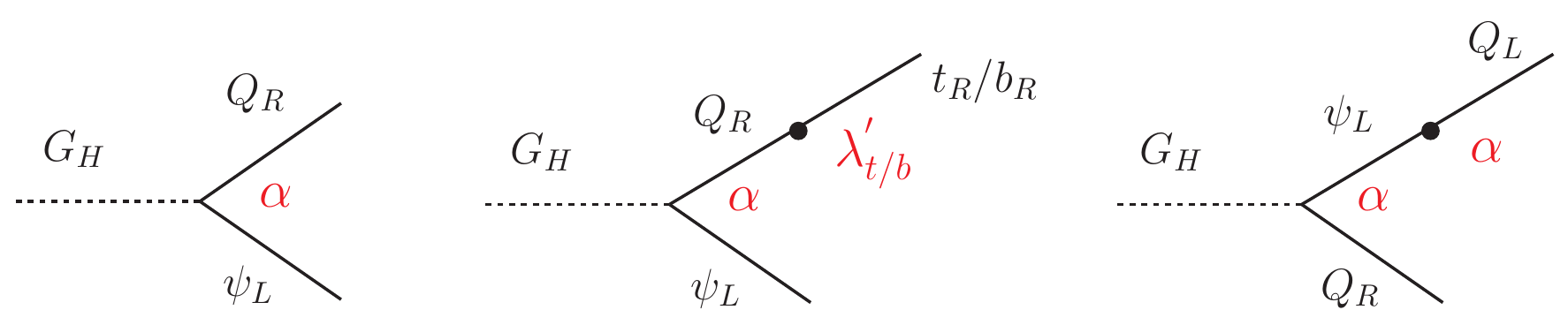}
\caption{ Feynman diagrams for the interactions of the color-octet scalar with $\psi Q$, $\psi t$ ($\psi b$), $QQ$; the $\psi$ states are first- and second-generation quarks while the $Q$ are heavy vector quarks.}
\label{fig:dia-GH-coupl}
\end{figure}

In order to identify the relevant signatures for $G_H$ searches at colliders, we need to determine the decay modes and calculate the corresponding rates. The decays of the new colored scalars include both flavor-diagonal and flavor-non-diagonal modes. The scalar's interactions with fermions arise from the Lagrangian term in Eq. (\ref{eq:Phi-term}).
In the case of the color-octet scalar, this directly generates the $G_H$ interactions with light and vector quarks:
\begin{equation}\label{eq:GH-psi-Q}
\frac{M}{u}\left( \vec{\bar{\psi}}_{L} \cdot \vec{\alpha} G^a_H T^a Q_{R} \right) + \text{H.c}.
\end{equation}
The mixing of $Q_R$ with $b_R$ and $t_R$, described by the matrices $\mathcal{D}_R$ and $\mathcal{U}_R$, in (\ref{matrix:DR4}), induces interactions of $G_H$ with light- and third-generation quarks
\begin{equation}\label{eq:GH-psi-b}
-\frac{M}{u}\left( \vec{\bar{\psi}}^d_{L} \cdot \vec{\alpha} G^a_H T^a \lambda^{'}_b b_{R} \right)-\frac{M}{u}\left( \vec{\bar{\psi}}^u_{L} \cdot \vec{\alpha} G^a_H T^a \lambda^{'}_t t_{R} \right) + \text{H.c.}~.
\end{equation}
Similarly, the mixing of $\psi_L$ with $Q_L$, described by the matrices $\mathcal{D}_L$ and $\mathcal{U}_L$,  in (\ref{matrix:DL4}), induces the interactions of $G_H$ with vector quarks
\begin{equation}\label{eq:GH-QQ}
\frac{M}{u}\left(|\alpha_1|^2+|\alpha_2|^2 \right) \bar{Q}_L G^a_H T^a  Q_{R}+ \text{H.c.} \ .
\end{equation}
Fig. \ref{fig:dia-GH-coupl} shows the Feynman diagrams, for $G_H$ interactions with ordinary quarks and vectorlike fields, including the relevant coupling factors.  Recall that we expect $\alpha_2$ is $\mathcal{O}(\lambda^2)$ and that $\alpha_1$ is $\mathcal{O}(\lambda^3)$.

At one loop, just as $G_H$ can be produced via gluon-gluon fusion, the octet scalar also decays to pairs of gluons as shown in Fig. \ref{fig:dia-Gamma0}. We denote the corresponding
rate by $\Gamma_0$ which we find to be, 
\begin{equation}
\Gamma_0 \left[G_H \to gg \right]=\frac{5 \alpha^2_s}{1536 \pi^3}\frac{\mu^2}{M_{G_H}}\left(\frac{\pi^2}{9}-1\right)^2~.
\end{equation}
Note that the rate is suppressed by a factor $\left(\frac{\pi^2}{9}-1\right)^2$ due to the destructive interference \cite{Bai:2010dj} between the triangle and the bubble loop in Fig. \ref{fig:dia-Gamma0}. 

The octet coupling to vector quarks in Eq. (\ref{eq:GH-QQ}) also contributes to $G_H$ decay into gluons, as shown in Fig. \ref{fig:dia-GammaQ}. The corresponding rate (neglecting interference with the scalar loop) is
\begin{equation}
\Gamma_Q \left[G_H \to gg \right]=\frac{5}{6}\frac{\alpha^2_s}{(16 \pi)^3}\frac{M^3_{G_H}}{u^2}\left(|\alpha_2|^2 +|\alpha_1|^2 \right)^2 \left|A\left(M^2_{G_H}/(4M^2)\right)\right|
\end{equation}
where A is the loop function \cite{Rizzo:1979mf}
\[
A(x)=\frac{2}{x^2}[x+(x-1)\text{Arcsin}^2\sqrt{x}] \qquad , \qquad \text{for} \ x\leq1 
\]
This rate is also suppressed by a factor $|\alpha_2|^4$ due to the flavor violating structure of our model, 
and therefore, in the following, we will neglect $\Gamma_Q$.

The rates for all relevant $G_H$ decay modes in our model read:
\begin{align}
\begin{split}\label{eq:G-rates}
& \Gamma\left[ G_H \to (\bar{d}_L b_R +\bar{b}_R d_L ) \right]= \vert \frac{\alpha_1}{\alpha_2}\vert^2 \ \Gamma \left[G_H \to (\bar{s}_L b_R +\bar{b}_R s_L )\right] = \frac{M_{G_H}}{16 \pi}\frac{M^2}{u^2}|\alpha_1 \lambda^{'}_b|^2\left(1-\frac{m^2_b}{M^2_{G_H}}\right)^2 \\
 & \Gamma\left[ G_H \to (\bar{u}_L t_R +\bar{t}_R u_L ) \right]= \vert \frac{\alpha_1}{\alpha_2}\vert^2 \   \Gamma \left[G_H \to (\bar{c}_L t_R +\bar{t}_R c_L )\right]  =\frac{M_{G_H}}{16 \pi}\frac{M^2}{u^2}|\alpha_1 \lambda^{'}_t|^2\left(1-\frac{m^2_t}{M^2_{G_H}}\right)^2 \\
&  \Gamma \left[G_H \to (\bar{d}_L Q^d_R +\bar{Q}^d_R d_L )\right]=\Gamma \left[G_H \to (\bar{u}_L Q^u_R +\bar{Q}^u_R u_L )\right]=\frac{M_{G_H}}{16 \pi}\frac{M^2}{u^2}|\alpha_1|^2\left(1-\frac{M^2}{M^2_{G_H}}\right)^2 \\
 &  \Gamma \left[G_H \to (\bar{s}_L Q^d_R +\bar{Q}^d_R s_L )\right]=\Gamma \left[G_H \to (\bar{c}_L Q^u_R +\bar{Q}^u_R c_L )\right]=\frac{M_{G_H}}{16 \pi}\frac{M^2}{u^2}|\alpha_2|^2\left(1-\frac{M^2}{M^2_{G_H}}\right)^2 \\
 & \Gamma \left[G_H \to \bar{Q}^d Q^d \right]= \Gamma \left[G_H \to \bar{Q}^u Q^u \right]=\frac{M_{G_H}}{16 \pi}\frac{M^2}{u^2}\left(|\alpha_2|^2 +|\alpha_1|^2 \right)^2\left(1-4\frac{M^2}{M^2_{G_H}}\right)^{3/2} \\
 & \Gamma_0 \left[G_H \to gg \right]=\frac{5 \alpha^2_s}{1536 \pi^3}\frac{\mu^2}{M_{G_H}}\left(\frac{\pi^2}{9}-1\right)^2\\
\end{split}
\end{align}
Depending on the octet mass, some of these decays will be below threshold and highly suppressed. In the next section we compute
the decay branching-fractions in different octet mass regions.

\subsection{Branching ratios} 

We will estimate the $G_H$ branching ratios (BR) in three different regions, depending on the $G_H$ mass: first $M_{G_H}<m_t$, then  $m_t<M_{G_H}<M$, and finally $M_{G_H}>M$.

\subsubsection{$M_{G_H}<m_t$}

When $G_H$ is lighter than the top quark,  the dominant decay modes are $G_H \to gg$ and $G_H \to s b$.  For the latter decay mode, applying the constraint in  Eq. (\ref{eq:bsgamma-lim}) to Eq. (\ref{eq:G-rates}) yields

\begin{equation}
\Gamma \left[G_H \to (\bar{s}_L b_R +\bar{b}_R s_L )\right]
\leq \frac{M_{G_H}}{16 \pi}\frac{m^2_b}{u^2}(0.0085)^2\left(1-\frac{m^2_b}{M^2_{G_H}}\right)^2
\label{eq:Gamma_sb}
\end{equation}
Likewise, by using the relationship $M^2_{\phi_I}=3\mu u$ of Eq. (\ref{eq:MphiI}), we can express the $\Gamma_0[G_H \to gg]$ rate from Eq. (\ref{eq:G-rates})   as
\begin{equation}
\Gamma_0 \left[G_H \to gg \right]
=\frac{5 \alpha^2_s}{(24 \pi)^3}\frac{M^4_{\phi_I}}{M_{G_H} u^2}\left(\frac{\pi^2}{9}-1\right)^2 \ .
\end{equation}

We therefore obtain the ratio:
\begin{equation}
\frac{\Gamma \left[G_H \to b_R s_L\right]}{\Gamma \left[G_H \to gg\right]}\lesssim 1 \cdot 10^3 \frac{m^2_b M^2_{G_H}}{M^4_{\phi_I}}\left(1-\frac{m^2_b}{M^2_{G_H}}\right)^2\  .
\label{eq:BR_bs}
\end{equation}
As a result, for $M_{\phi_I}\sim M_{G_H}$,  we get $BR[G_H \to b_R s_L]\simeq 1$ if we set $|\alpha_2\lambda^{'}_b/\beta_b|$ to the maximum value allowed by $b\to s \gamma$.   On the other hand, for $M_{\phi_I}\gtrsim 2 M_{G_H}$, the decay to gluons becomes relevant.  

For $M_{G_H}\gtrsim 120$ GeV the decay of the octet into an off-shell top and a charm, leading to a $Wbc$ final state becomes important. This is similar to the 3-body decay computed in Eqs (3.13)-(3.17)
of \cite{Dobrescu:2011px}. Adapting these formulas to our model, we find a value of 0.13 for the $G_H \to t^{*}c \to Wbc$ decay branching-ratio at $M_{G_H}=120$ GeV. The $BR[G_H \to t^{*}c]$ increases for heavier octets and the $G_H \to t^{*}c\to Wbc$ decay mode becomes dominant for $M_{G_H}\gtrsim 150$ GeV -- in particular $BR[G_H \to t^{*}c]=0.64$ for $M_{G_H}= 150$ GeV.

The left pane of 
Fig. \ref{fig:width-GH} shows the $G_H$ total decay width, which is of the order $10^{-8}$ GeV ($c\tau \sim 20 nm$), in this mass region, for $u=1$ TeV and for three values of $M_{\phi_I}/ M_{G_H}$. \\

\subsubsection{$m_t<M_{G_H}<M$}

Next, we estimate the $G_H$ branching ratios in the mass region when the color-octet scalar is heavier than the top quark but lighter than the vector quarks. Given that $|\lambda^{'}_b| \ll |\lambda^{'}_t|$ and $|\alpha_1|\ll |\alpha_2|$, we will neglect $G_H$ decays into bottom quarks and into first-generation quarks in the following.
In the limit $|\lambda^{'}_b| \ll |\lambda^{'}_t|$ we can also use the relationship $|\alpha_2 \lambda^{'}_t|/\beta_t=A\lambda^2=0.0415$ and the estimate $\beta_t\simeq m_t/M$ to write:
\begin{align}
\begin{split}\label{eq:G-rates-2}
 & \Gamma(G_H \to t c) \equiv \Gamma \left[G_H \to (\bar{c}_L t_R +\bar{t}_R c_L )\right]=\frac{M_{G_H}}{16 \pi}(A\lambda^2)^2\frac{m^2_t}{u^2}\left(1-\frac{m^2_t}{M^2_{G_H}}\right)^2 
 \end{split}
\end{align}
Note that $G_H$ decays equally to $\bar{c}_L t_R$  {\it and} $\bar{t}_R c_L$ -- to both top quarks {\it and}
anti-top quarks. This property leads to a same-sign dilepton signal for the {\it pair} production of the color-octet scalars, as discussed further in the next section.

\begin{figure}
\includegraphics[width=.48\textwidth]{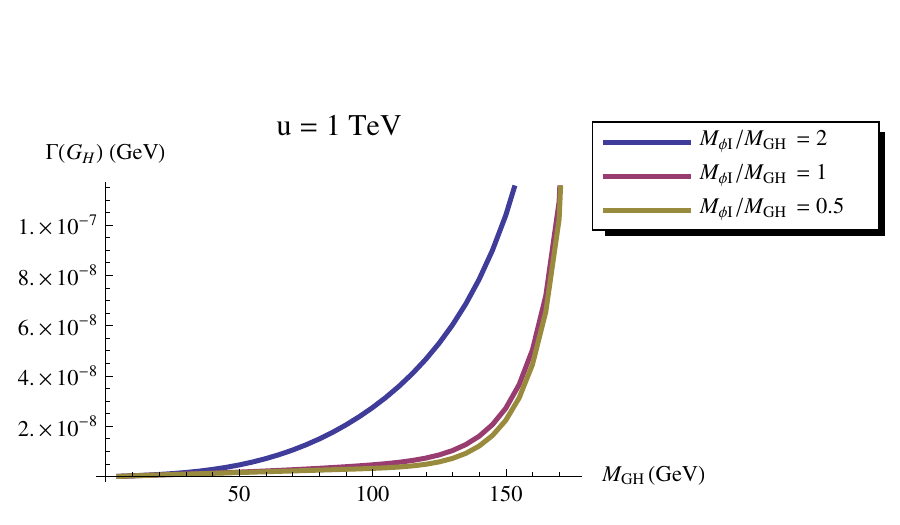}
\includegraphics[width=.48\textwidth]{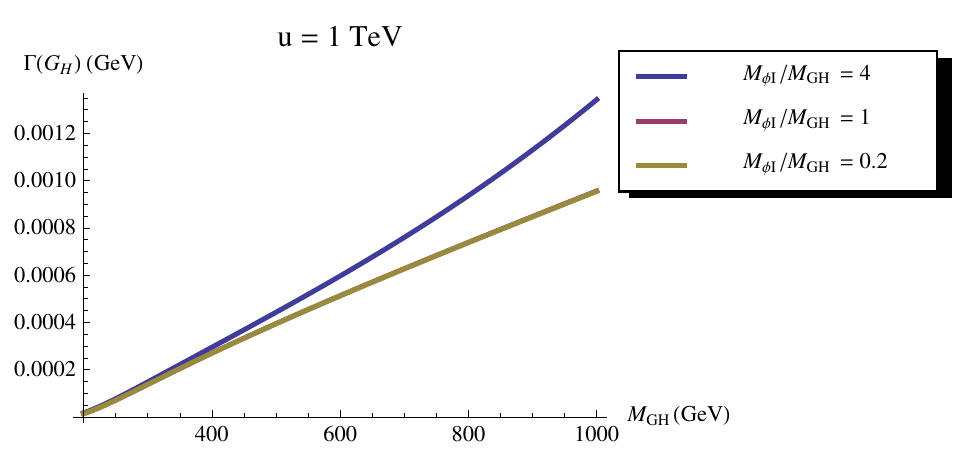}
\caption{$G_H$ total decay width as function of $M_{G_H}$ for three different $M_{\phi_I}/M_{G_H}$ values (see the key) and for $u=1$ TeV.  Left Plot: $M_{G_H}<m_t$. Right Plot: $m_t<M_{G_H}<M$. The width scales as $(1\ \text{TeV}/u)^2$ with $u$.}
\label{fig:width-GH}
\end{figure}

\begin{figure}
\includegraphics[width=.46\textwidth]{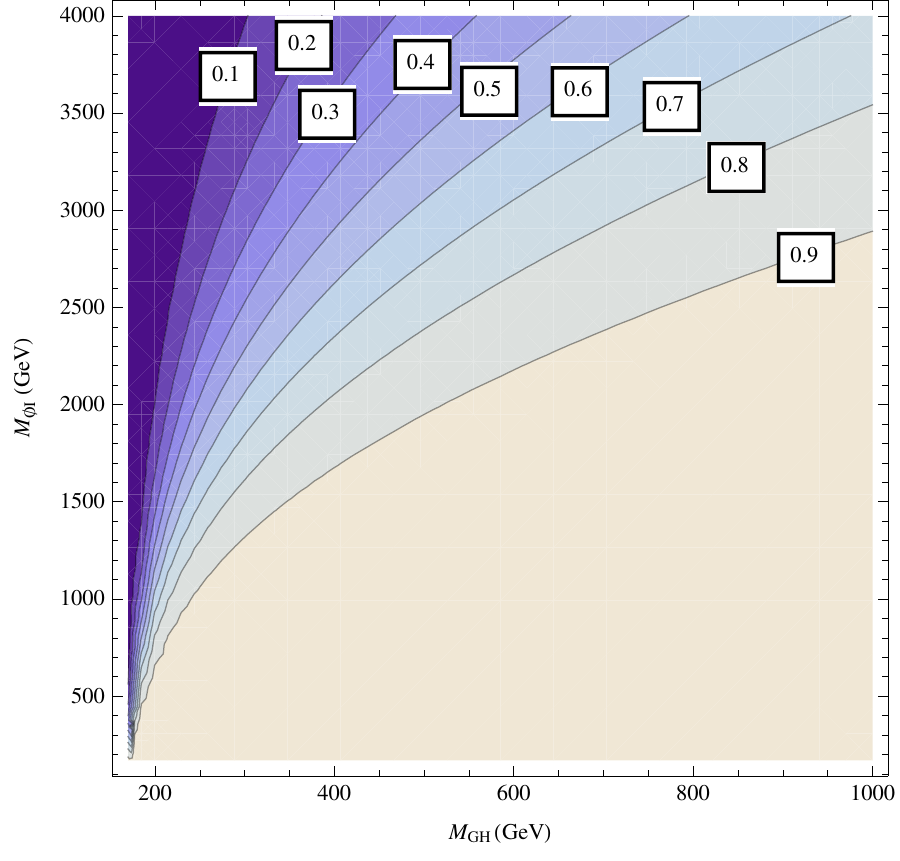}
\caption{Contour plot for $BR\left[G_H \to tc \right]$ in the plane $(M_{G_H}, M_{\phi_I})$.  }
\label{fig:contour-MGH-MphiI}
\end{figure}

We find
\begin{equation}
\frac{\Gamma\left[G_H \to tc \right]}{\Gamma\left[G_H \to gg \right]}
\simeq 2.2 \cdot 10^3 \, \frac{M^2_{G_H} m^2_t}{M^4_{\phi_I}}\left(1-\frac{m^2_t}{M^2_{G_H}}\right)^2~.
\label{eq:ratio_tc-gg}
\end{equation}
As shown in Fig. \ref{fig:contour-MGH-MphiI}, we find that $BR\left[G_H \to tc \right]\simeq 1$, except in the small region of parameter space where $M_{\phi_I}\gg 4 M_{G_H}$. The right pane of  Fig. \ref{fig:width-GH} displays the octet scalar's width in this intermediate mass range for $u=1$ TeV and  three values of $M_{\phi_I}/ M_{G_H}$; the width is of the order of $10^{-4}$ GeV and scales as $(1\ \text{TeV}/u)^2$.

\subsubsection{$M_{G_H}>M$}

\begin{figure}
\includegraphics[width=.45\textwidth]{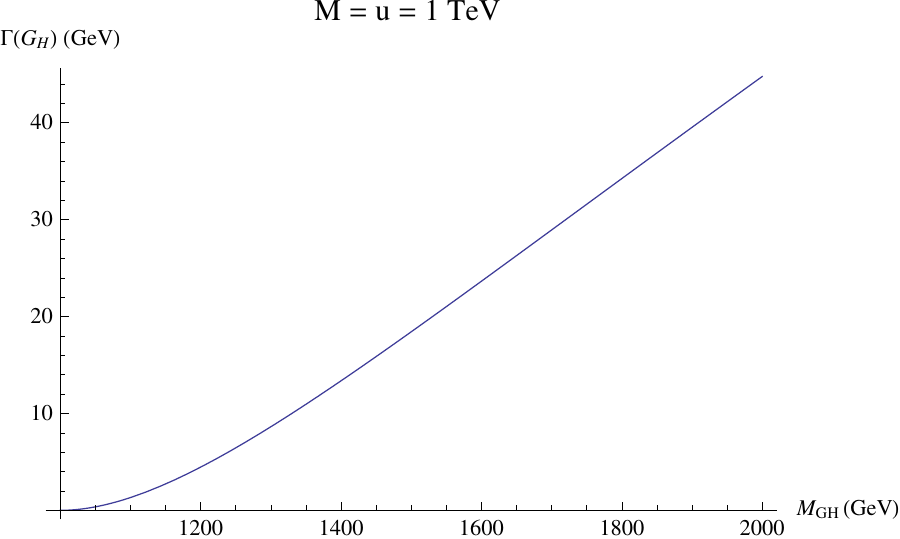}
\caption{$G_H$ total decay width as function of $M_{G_H}$ for $M=u=1$ TeV. The width scales as $(M/u)^2$.}
\label{fig:width-Qpsi}
\end{figure}

Finally, we consider the case of heavy color-octet scalars, with $M_{G_H}>M$. In this case, 
$G_H$ is above the threshold for the decay into $Q^d_R s_L$ and $Q^u_R c_L$, and we have
 \begin{equation}
\frac{\Gamma \left[G_H \to Q^u_R c_L\right]}{\Gamma \left[G_H \to t_R c_L\right]} \simeq |\lambda^{'}_t|^{-2} \left(1-\frac{M^2}{M^2_{G_H}}\right)^2 \sim 25 \left(1-\frac{M^2}{M^2_{G_H}}\right)^2
\label{eq:BR_Qpsi}
\end{equation}
where we have used the fact that $|\lambda^{'}_t|$ is of the order $\beta_t\sim m_t/M \sim 0.2$ for $M\sim 1$ TeV.
In this mass region $BR[G_H \to  Q^u_R c_L \ , \ Q^d_R s_L] \simeq 1$. Fig. \ref{fig:width-Qpsi} shows the $G_H$ total decay width, which is of the order of 10 GeV, for $M=u=1$ TeV.

\subsection{Limits on the octet mass}

Given the information on the $G_H$ decay patterns, we can use existing collider data to place bounds on the octet scalar's mass.

\subsubsection{$M_{G_H}< m_t$}

In the mass region $M_{G_H}<m_t$, the $G_H$ decays only into jets\footnote{Except in the case $M_{G_H}\gtrsim 150$ GeV, where $G_H \to t^{*}c \to Wbc$ becomes the dominant decay mode.}: either gluons or b-jet plus jet. The recent Tevatron search for pair produced dijet resonances \cite{Aaltonen:2013hya} therefore
excludes the region $50{\,\rm  GeV} < M_{G_H} < 125 {\, \rm GeV}$.\footnote{The Tevatron search considers the pair production of colored octets as in Ref. \cite{Bai:2010dj}. The only difference with our model is the possibility for the octet to decay not only into gluons but also into a b-jet plus a light-jet. However, since no b-tagging is applied in the analysis we can directly apply the limit above to our case}

\subsubsection{$m_t<M_{G_H}<M$}

In the mass region $m_t<M_{G_H}<M$, the octet decays into $tc$, $t\bar{c}$ or $\bar{t}c$, unless $M_{\phi_I}\gg 4 M_{G_H}$. The pair-production of octets in this mass region, considering the leptonic decays of the two tops from the octets, can therefore lead to same-sign-dilepton (SSD) final states. CMS searches fix 95\% C.L. upper limits on the number of single-sign dilepton events \cite{Chatrchyan:2012paa}. The most recent CMS study is for an integrated luminosity of $10.5$ fb$^{-1}$ collected at the 8 TeV LHC \cite{SUS-12-029}. 

\begin{figure}
\includegraphics[width=.95\textwidth]{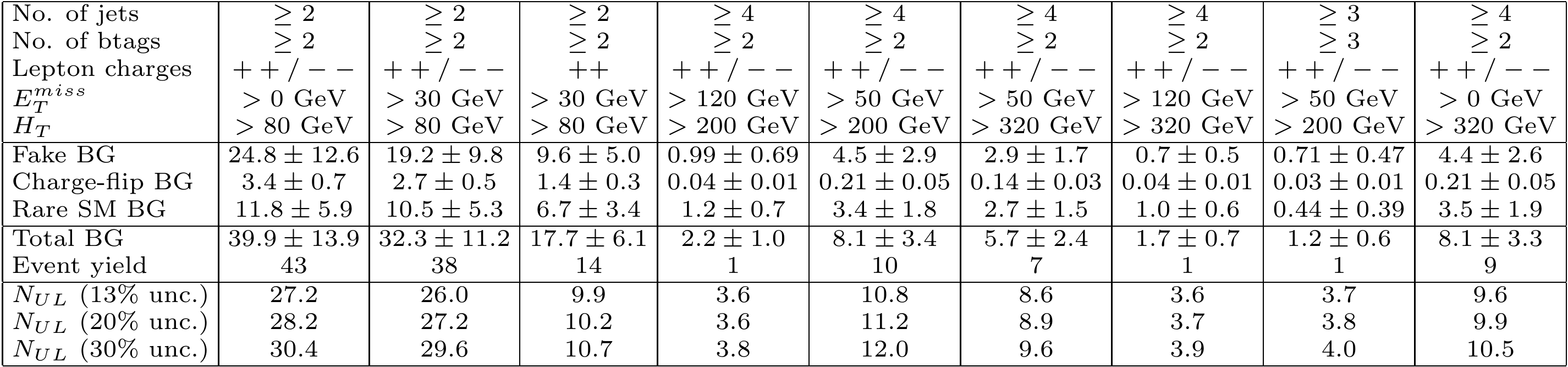}
\caption{Table 2 from \cite{SUS-12-029}. Results for different signal regions (SR), starting with SR0 at left and proceeding all the way to SR8 in the right-most column. For each signal region, the table shows the most distinctive kinematic requirements, the prediction for the three background (BG) components, the total background and the number of observed events; the last three rows show the 95\% CL upper limits corresponding to three possible values of the signal efficiency uncertainty. The number of jets on the first line of the table includes both tagged and untagged jets.}
\label{fig:table-sr}
\end{figure}

In order to compare these limits with the prediction from our model, we have followed the procedure explained in sec. 7 of \cite{Chatrchyan:2012sa}. In particular, we have used  MADGRAPHv5 to simulate the signal events 
\[
pp \to G_H G_H \to l l ^{(++/--)} + 2b + 2j + E^{miss}_T
\]
with $l=e,\mu$, assuming a 100\% $G_H$ decay BR into $tc$. We have then applied the same kinematic cuts on jets and leptons used in the CMS analysis \cite{SUS-12-029} and we have considered the cuts on the missing transverse energy and on the $H_T$ variable (defined as the scalar sum on the $p_T$ of all the final jets, including b-jets) corresponding to those applied in the different signal regions (SR) listed on Table 2 of \cite{SUS-12-029} (reported in Fig. \ref{fig:table-sr}).  We have considered all of the different signal samples defined there, except for the selection SR7 which requires the tagging of at least 3 b-jets. Using the results of \cite{SUS-12-029}, the calculated cross sections have been corrected for trigger efficiencies for $ee$, $e\mu$, $\mu\mu$ detection, lepton selection efficiencies, b-tagging efficiencies and for the efficiencies on the $H_T$ and $E^{miss}_T$ cuts.

\begin{figure}
\includegraphics[width=.55\textwidth]{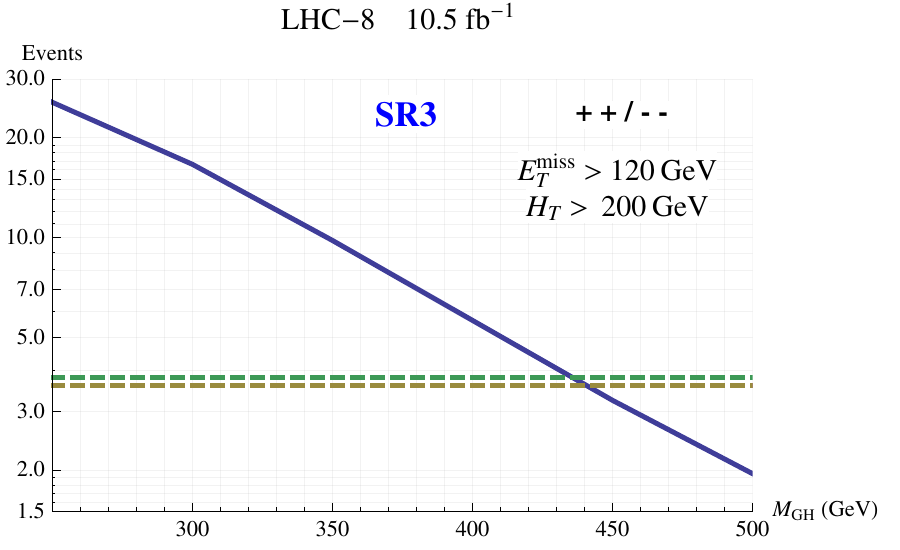}
\caption{Number of events, at $10.5$ fb$^{-1}$ at the 8 TeV LHC, with same-sign-dileptons passing the so-called SR3 selection in \cite{SUS-12-029} as a function of the octet mass in our model (solid curve) compared with the CMS \cite{SUS-12-029} 95$\%$ C.L.  upper limits (dashed horizontal lines) for different signal efficiency uncertainties.  The uncertainty is either 13\% or 20\% for the lower line and 30\% for the upper line; as shown in Fig. \ref{fig:table-sr}, the two lower uncertainty values happen to yield the same limit for case SR3.}
\label{fig:sr3}
\end{figure}

\begin{figure}
\includegraphics[width=.45\textwidth]{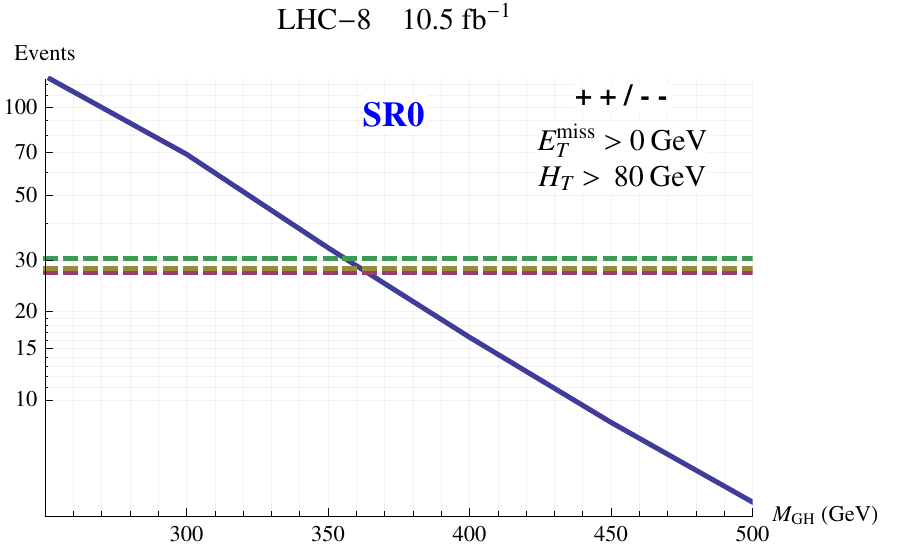}
\includegraphics[width=.45\textwidth]{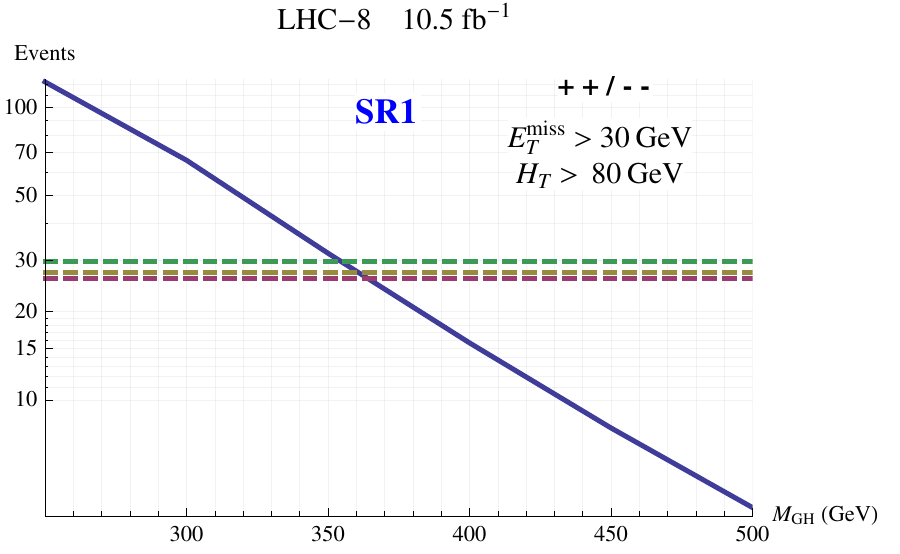}
\includegraphics[width=.45\textwidth]{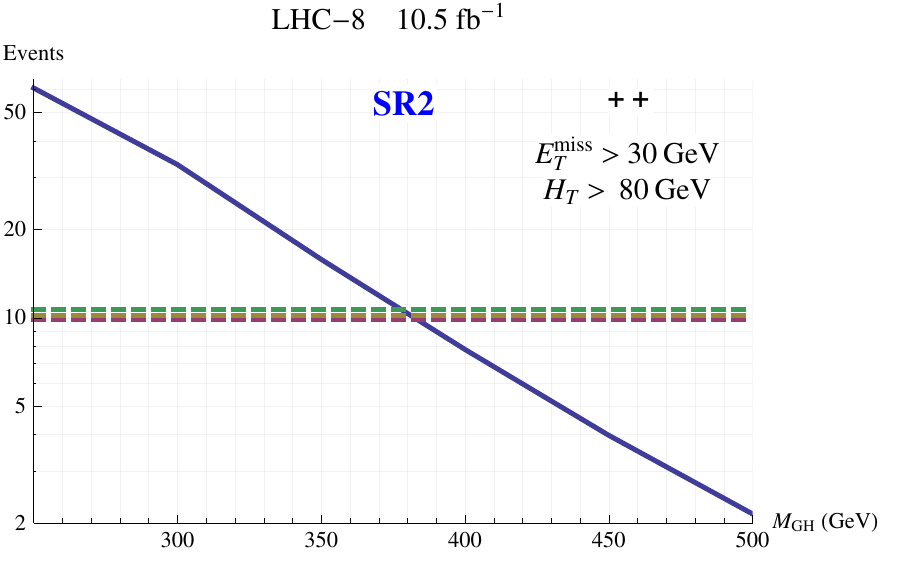}
\includegraphics[width=.45\textwidth]{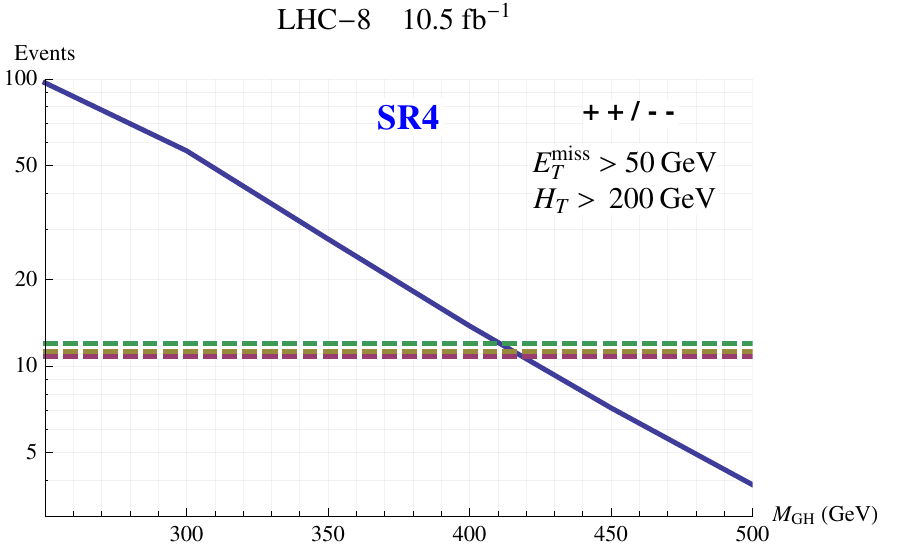}
\includegraphics[width=.45\textwidth]{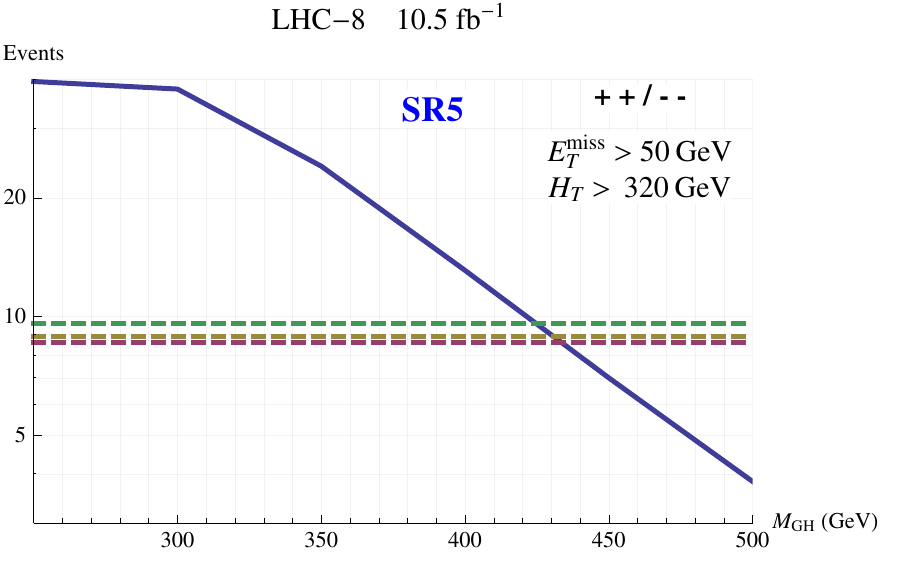}
\includegraphics[width=.45\textwidth]{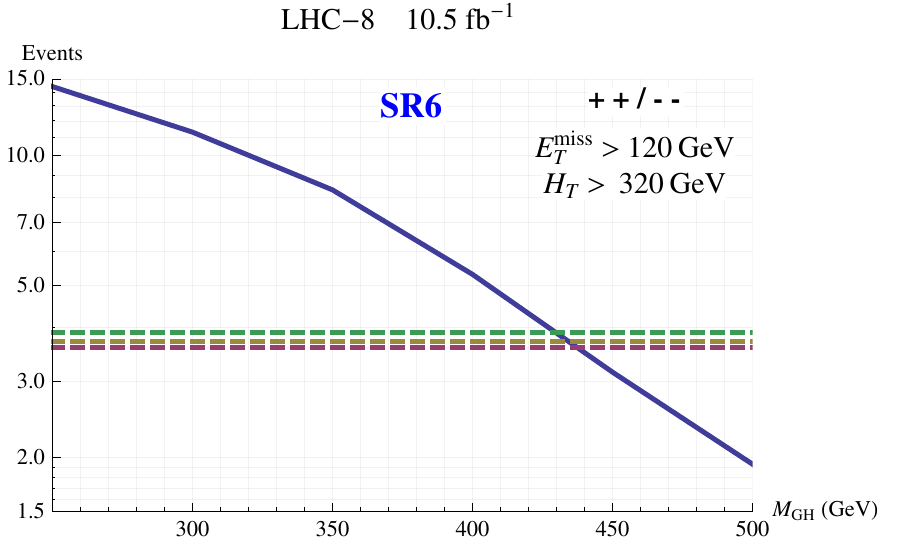}
\includegraphics[width=.45\textwidth]{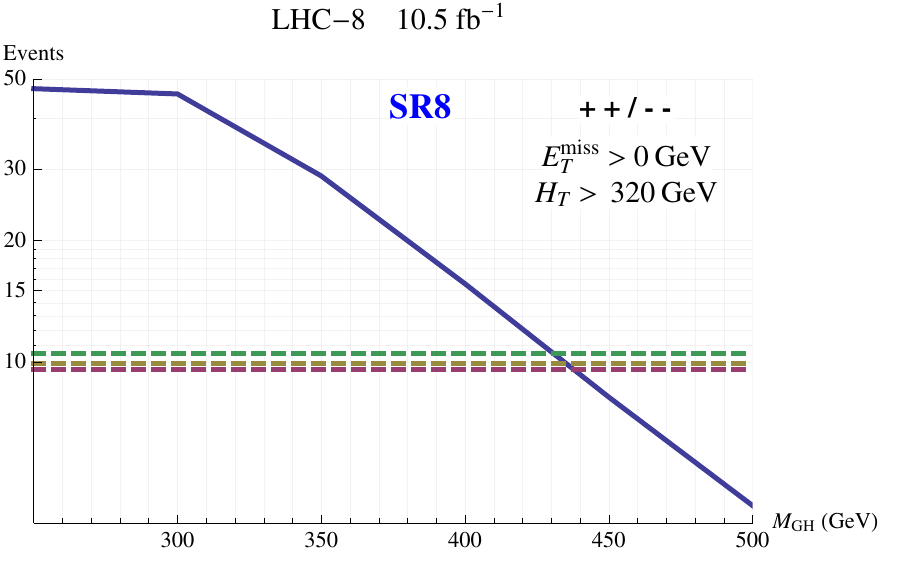}
\caption{Number of events, at $10.5$ fb$^{-1}$ at the 8 TeV LHC, with same-sign-dileptons passing the different SR selections in \cite{SUS-12-029} as a function of the octet mass in our model (solid) compared with the CMS \cite{SUS-12-029} upper limits at $13\%$, $20\%$ and $30\%$ uncertainty in the signal efficiency (horizontal dashed lines from lowest to highest).}
\label{fig:ssd-tot}
\end{figure}

The final number of events at $10.5$ fb$^{-1}$ for different $G_H$ masses is then compared to the CMS limits. We find that the SR3 sample, with the cuts $E^{miss}_T>120$ GeV and $H_T>200$ GeV, gives the strongest constraint on the octet mass, namely:
\begin{equation}\label{eq:MGH-limit-ssd}
M_{G_H} \gtrsim 440 \, \text{GeV}\,\qquad (M_{G_H} > M_{\phi_I}/4)~.
\end{equation}
at 95\% CL. Fig. \ref{fig:sr3} shows the number of events with same-sign-dileptons passing the selection SR3 as a function of the octet mass in our model and the CMS observed 95$\%$ confidence level upper limits calculated under three different assumptions for the signal efficiency uncertainty, namely  $13\%$, $20\%$ and $30\%$. We also show (in Fig. \ref{fig:ssd-tot}) the model predictions and the CMS limits for the other samples, which give slightly milder bounds on $M_{G_H}$ than that in (\ref{eq:MGH-limit-ssd}); we make no attempt to combine these limits.

\begin{figure}
\includegraphics[width=.48\textwidth]{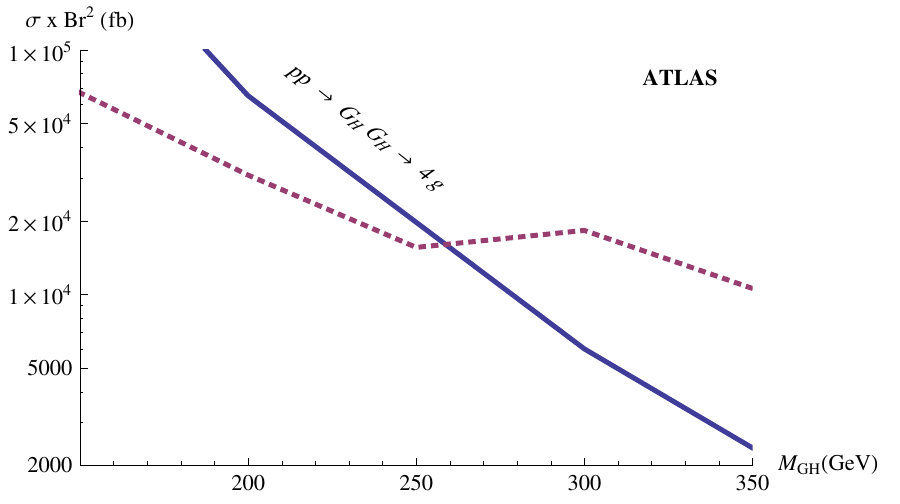}
\includegraphics[width=.48\textwidth]{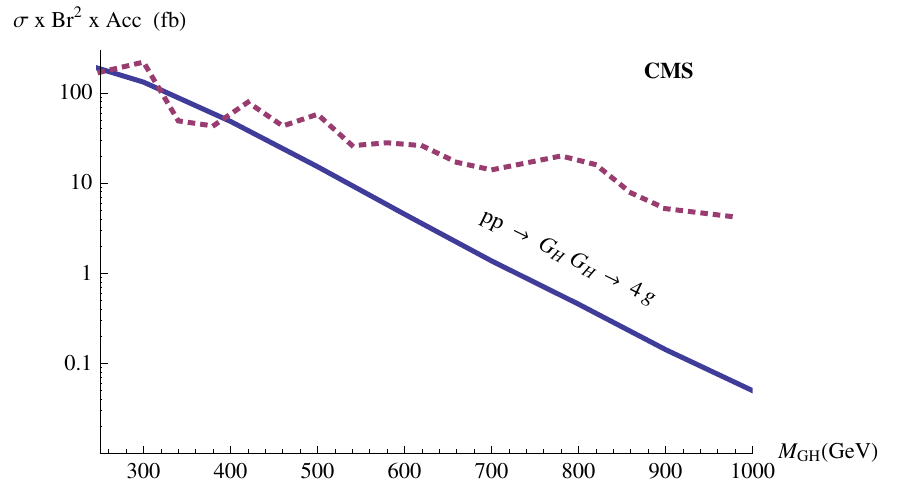}
\caption{ $pp\to G_H G_H$, assuming a 100\% decay branching-ratio of the octet to gluon pairs. Left Plot: cross section (solid) and the observed ATLAS limits from \cite{ATLAS:2012ds} (dashed). Right Plot: cross section times acceptances (solid) and the observed CMS limits from \cite{Chatrchyan:2013izb} (dashed). }
\label{fig:cms-dijet-pair}
\end{figure}

For 
$M_{\phi_I}\gg 4 M_{G_H}$
the octet decays into gluons become relevant. ATLAS \cite{ATLAS:2012ds} and CMS \cite{Chatrchyan:2013izb} have recently performed a search for pair-produced dijet resonances at the 7 TeV LHC. ATLAS presents upper bounds on colored scalar pair production cross-section times branching ratio to gluons, as a function of the scalar mass, from which we can extract limits on the $G_H$ mass. Fig. \ref{fig:cms-dijet-pair} (left plot) shows the $pp\to G_H G_H$ cross section, assuming a 100\% decay branching-ratio of the octet to gluon pairs, and the observed ATLAS limits from \cite{ATLAS:2012ds}.
CMS has also considered the pair production of colorons, taking as reference the model in \cite{Dobrescu:2007yp} and assuming that the coloron decays entirely into jets. We have applied the same acceptances for the coloron, quoted in \cite{Chatrchyan:2013izb}, to the color-octet scalar $G_H$. The resulting values of cross section times acceptances for $pp\to G_H G_H$, assuming a 100\% decay branching-ratio of the octet to gluon pairs,  are shown, together with the observed CMS limits in Fig. \ref{fig:cms-dijet-pair} (right plot). 
We see that, in this case $M_{\phi_I}\gg 4 M_{G_H}$, the ATLAS and CMS analyses exclude the mass regions:
\begin{align}
\begin{split}
& \text{Excluded regions:}  \qquad  (M_{G_H}\ll M_{\phi_I}/4) \\[0.4cm]
& [ATLAS] \qquad 150 \, \text{GeV} < M_{G_H} < 260 \, \text{GeV} \\
& [CMS] \qquad 320 \, \text{GeV} < M_{G_H} < 390 \, \text{GeV} 
\end{split}
\end{align} 

\begin{figure}
\includegraphics[width=.48\textwidth]{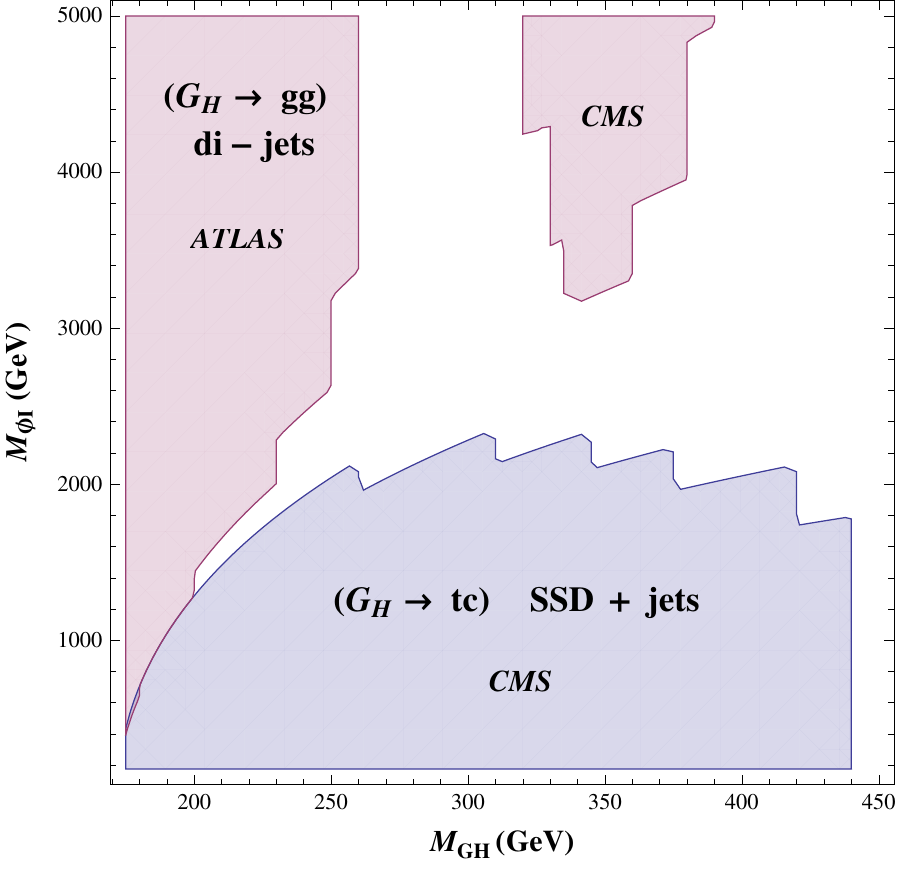}
\caption{Exclusion regions in the plane $(M_{G_H}, M_{\phi_I})$. The lower blue region is excluded by the CMS search for new physics in same-sign-dileptons channels \cite{SUS-12-029}. The upper red regions are excluded by the ATLAS \cite{ATLAS:2012ds} and CMS \cite{Chatrchyan:2013izb} searches for pair-produced dijet resonances.}
\label{fig:octet-excl}
\end{figure}

The plot in Fig. \ref{fig:octet-excl} shows the exclusion regions in the plane $(M_{G_H}, M_{\phi_I})$, obtained by combining the limits derived from the CMS analysis in same-sign-dileptons channels \cite{SUS-12-029}, where the octet decays into $tc$, with those derived from the ATLAS \cite{ATLAS:2012ds} and CMS \cite{Chatrchyan:2013izb} searches for pair-produced dijet resonances, where $G_H$ decays into a gluon pair.


\subsubsection{$M_{G_H}>M$}

When $G_H$ is above the threshold for decaying into vector quarks,
$G_H \to Q^{u,d}_R \psi^{u,d}_L$ are the dominant decay modes. An ATLAS analysis \cite{atlas-conf-2012-130} places the constraint on the vector-quark mass $M\gtrsim 640$ GeV. The heavy vectors decay to $Vt_R$, with $V$ representing $h$, $W$ or $Z$ (\ref{eq:BR-Q}). The pair production of $G_H$ in this region would therefore yield  spectacular signatures $VVtt+jets$.  Thus far, there are no constraints available on these final states.

\ \

Table \ref{tab:GH-pheno} summarizes the above results, showing the dominant decay modes and the most promising signatures for the discovery of the octet scalars in each mass range.

\begin{table}
\begin{tabular}{|c|c|c|c|}
\hline 
{\bf Mass Regions} & {\bf Dominant Decays} & {\bf Signatures} & {\bf Excluded Regions}\\[0.1cm]
\hline
$M_{G_H}<m_t$ & $G_H \to b_R s_L \, , qq $ & dijet pairs & [50 GeV, 125 GeV] (CDF)\\[0.1cm]
$m_t<M_{G_H}<M$ & $G_H \to t_R\bar{c}_L +\bar{t}_R c_L $ & SSD & [$m_t$, 440 GeV] (CMS)\\[0.1cm]
$M_{G_H}>M$ & $G_H \to Q_R\psi_L $ & $VVtt+jets\to$ SSD & \\
\hline
\end{tabular}
\caption{Dominant decay modes and most promising signatures for the discovery of the color-octet scalar, depending on its mass. We also show the mass regions already excluded by CDF \cite{Aaltonen:2013hya} and CMS \cite{SUS-12-029} searches for dijets and same-sign dileptons. }
\label{tab:GH-pheno}
\end{table}

\section{Color-Singlet Pseudoscalar Phenomenology}
\label{Sec:singlet}

In this section, we describe the relevant production modes and dominant decays of the color-singlet pseudoscalar $\phi_I$.

\subsection{$\phi_I$ Production}

Due to an intrinsic parity symmetry of the scalar potential, interactions of one color-singlet pseudoscalar with a pair of color-octets, which one might have expected to arise from the $\mu$ cubic term in (\ref{eq:potential}), are forbidden. This implies that the only mechanism for direct single production of $\phi_I$  is gluon-gluon fusion via loop-exchange of vectorlike quarks through the interactions
\begin{equation}
\frac{i}{\sqrt{6}} \frac{M}{u} |\alpha_2|^2 \bar{Q}_L \phi_I Q_R \ .
\end{equation}
which are suppressed by a factor of $|\alpha_2|^2$.
The diagram for this production mechanism is analogous to that in Fig. \ref{fig:dia-GammaQ} for the single production of $G_H$ via loop exchange of heavy vector-like $Q$ fermions. The corresponding rates are very small, of the order of attobarns at the 14 TeV LHC and, as such, not observable at the LHC \footnote{Note that the rate of pseudoscalar production in our model is different from that considered in other studies in the literature.  For example in Ref. \cite{Gresham:2007ri}, the single production of colored scalars induced by top-loops (negligible in our model) is sizable and gives the main contribution to the single-production channel.}. 

Nevertheless, as we will discuss in Sec. \ref{Sec:coloron}, there is one promising channel to directly observe the color-singlet pseudoscalar at the 14 TeV LHC: production of $\phi_I$ in association with $G_H$, from the decay of a coloron.

\subsection{$\phi_I$ decay}

The interactions of the color-singlet pseudoscalar with fermions, as for those of the color-octet, are generated from the term in (\ref{eq:Phi-term}) and can be obtained from those of the color-octet by substituting $T^{a} \to i\mathcal{I}/\sqrt{6}$. The corresponding diagrams are analogous to those for $G_H$, shown in Fig. \ref{fig:dia-GH-coupl}. The rates for the relevant $\phi_I$ decay modes are:
\begin{align}
\begin{split}
& \Gamma \left[\phi_I \to (\bar{s}_L b_R +\bar{b}_R s_L )\right]=\frac{M_{\phi_I}}{16 \pi}\frac{M^2}{u^2}|\alpha_2 \lambda^{'}_b|^2\left(1-\frac{m^2_b}{M^2_{G_H}}\right)^2 \leq \frac{M_{\phi_I}}{16 \pi}\frac{m^2_b}{u^2}(0.0085)^2\left(1-\frac{m^2_b}{M^2_{G_H}}\right)^2\\
& \Gamma \left[\phi_I \to (\bar{c}_L t_R +\bar{t}_R c_L )\right]=\frac{M_{\phi_I}}{16 \pi}\frac{M^2}{u^2}|\alpha_2 \lambda^{'}_t|^2\left(1-\frac{m^2_t}{M^2_{\phi_I}}\right)^2 \\
 \label{eq:Gamma_sb_PhiI}
\end{split}
\end{align}
where we have used the constraint in (\ref{eq:bsgamma-lim}).
As for the octet, the decay of $\phi_I$ into an off-shell top and a charm is relevant in the $\phi_I$ mass region from 120 GeV to $m_t$. We find that $BR[\phi_I \to t^{*}c]=0.2\, (0.6)$ for $M_{\phi_I}=120\, (140)$ GeV.

\begin{figure}
\includegraphics[width=.45\textwidth]{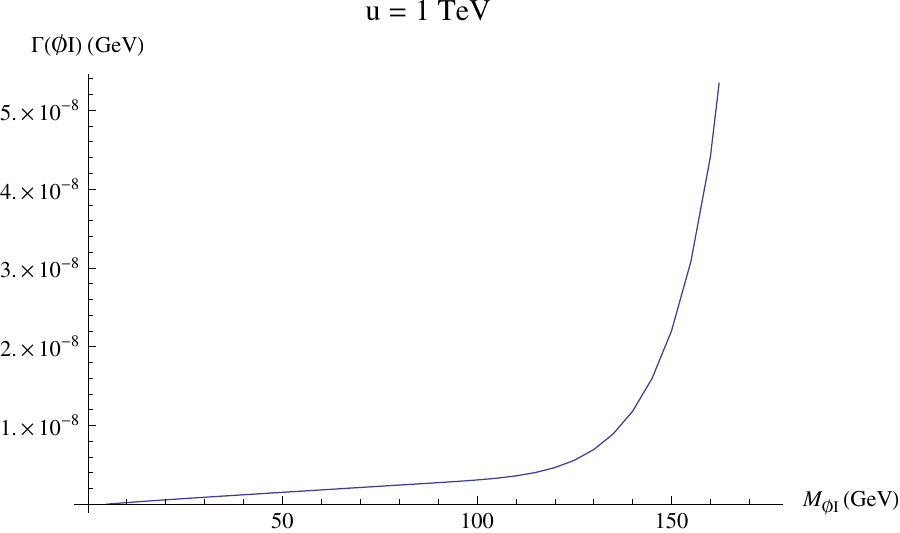}
\includegraphics[width=.45\textwidth]{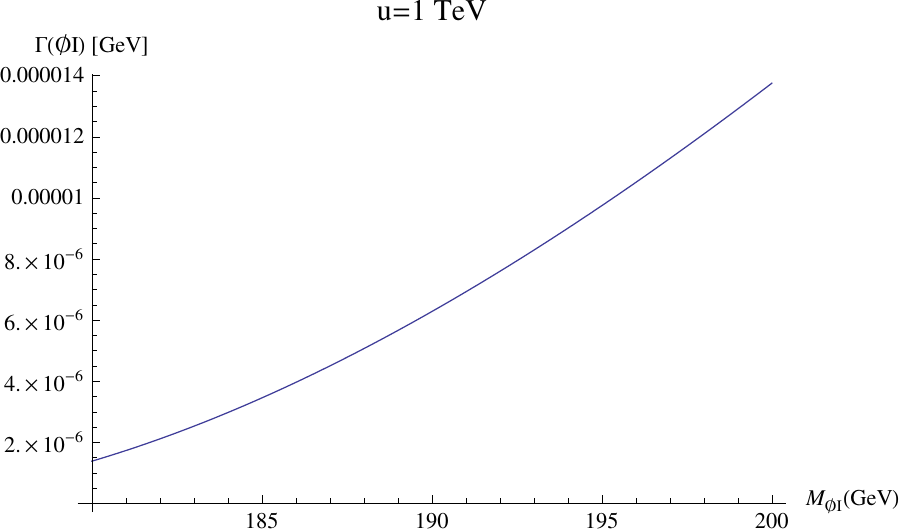}
\caption{Total decay width for $\phi_I$ as a function of $M_{\phi_I}$ for $u=1$ TeV. The width scales as $(1\ \text{TeV}/u)^2$ with $u$.  The left pane applies for $M_{\phi_I} < m_t$; the right pane is for scalars heavier than the top quark.}
\label{fig:width}
\end{figure}

Fig. \ref{fig:width} shows the $\phi_I$ total decay width as a function of $M_{\phi_I}$ for $u=1$ TeV. The width scales as $(1\ \text{TeV}/u)^2$ with $u$. For $u=1$ TeV and $M_{\phi_I}<m_t$ the width is of the order of $10^{-9}$ GeV ($c\tau \sim 200$ nm), for $|\alpha_2\lambda^{'}_b/\beta_b|$ close to the maximum allowed value from $b\to s \gamma$. For $u=1$ TeV and $M_{\phi_I}>m_t$ the width is of the order of $10^{-5}$ GeV.

\section{Coloron Phenomenology with Light Colored Scalars}
\label{Sec:coloron}

Ref. \cite{Chivukula:2013kw} presents a study of the coloron phenomenology in our model, in the case where the new colored scalars are heavier than the coloron gauge bosons. In this section we will discuss the coloron phenomenology in the cases of color-ocet and color-singlet scalars lighter than  the colorons. 

A coloron has no tree-level three-point coupling to gluon pairs and is produced at colliders through its interactions with light quarks. The rates for the relevant coloron decays read:
\begin{align}
\begin{split}\label{eq:BR-coloron}
&\Gamma(C\to jj)=\frac{1}{6\pi}g^2_S M_C \tan^2\omega \\
&\Gamma(C\to t\bar{t})=\frac{1}{24\pi} g^2_S M_C \cot^2\omega\sqrt{1-4\frac{m^2_t} {M^2_C}}\left(1+2\frac{m^2_t}{M^2_C}\right) \\
&\Gamma(C\to b\bar{b})=\frac{1}{24\pi} g^2_S M_C \cot^2\omega \\
\end{split} 
\end{align}
where $j=u,d,c,s$. As we found in \cite{Chivukula:2013kw}, the ATLAS search for new resonances decaying to dijets, based on 8 TeV data \cite{atlas-conf-2012-148}, gives a lower bound on the coloron mass that ranges from $M_C > 2.4$ TeV for $\cot\omega \approx 2.5$, when the coloron couples more strongly to third generation quarks, all the way to $M_C > 4.3$ TeV for $\cot\omega \approx 0.5$, when the coloron couples mainly to light quark generations. In deriving these limits, we assumed that corrections to the coloron branching ratios coming from the coloron's interactions with color-octet and color-singlet scalars were negligible.

\begin{figure}
\includegraphics[width=.45\textwidth]{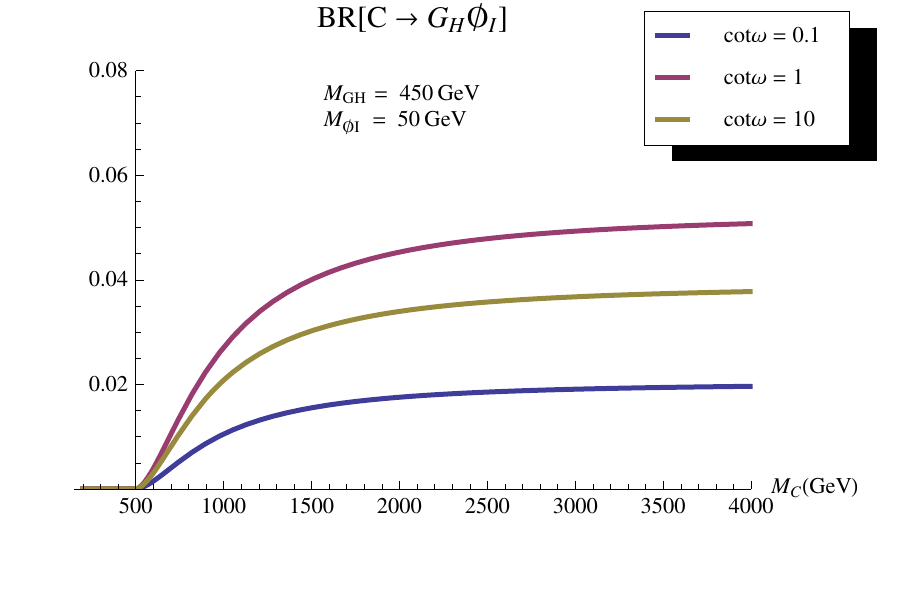}
\includegraphics[width=.45\textwidth]{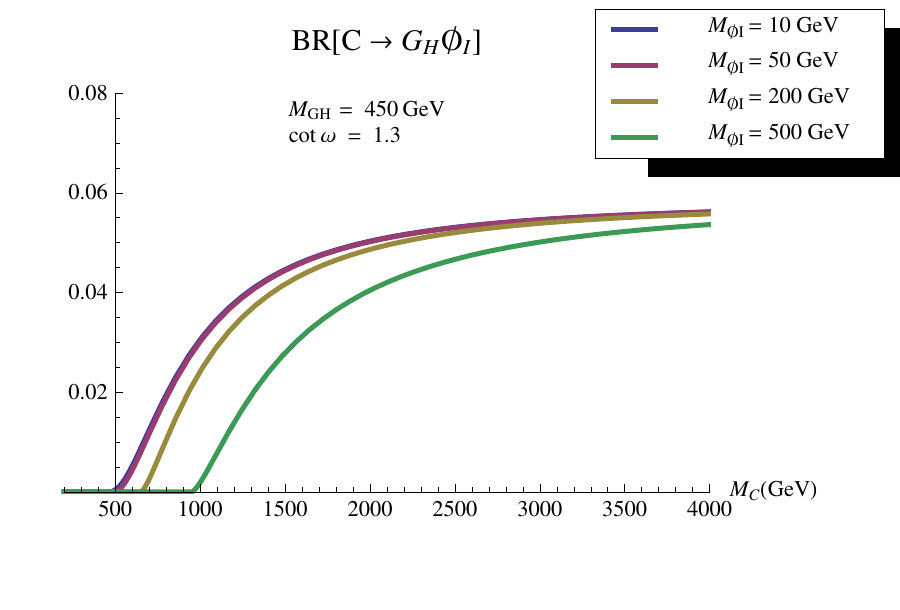}
\caption{The coloron decay branching ratio into $G_H \phi_I$ as a function of $M_C$, for different $\cot\omega$ values (left pane) and  for different pseudoscalar masses (right pane).  }
\label{fig:coloron-scalars}
\end{figure}

If the color-octet $G_H$ and the color-singlet pseudoscalar $\phi_I$ are lighter than the colorons, more precisely for $M_C>2 M_{G_H}$ and $M_C>( M_{G_H}+M_{\phi_I)}$, colorons can decay at tree level into $G_H G_H$ or into $G_H \phi_I$ respectively. The rates for these decays are:
\begin{align}
\begin{split}
& \Gamma \left[C \to G_H G_H\right]=\alpha_s\frac{\left(\cot^2\omega-1\right)^2}{64\cot^2\omega}M_C \left(1-4\frac{M^2_{G_H}}{M^2_C}\right)^{3/2} \\
& \Gamma \left[C \to G_H \phi_I\right]=\alpha_s\frac{\left(\cot^2\omega+1\right)^2}{72\cot^2\omega}M_C \left(1-2\frac{M^2_{\phi_I}+M^2_{G_H}}{M^2_C}+\frac{\left(M^2_{\phi_I}-M^2_{G_H}\right)^2}{M^4_C}\right)^{3/2} 
 \label{eq:Gamma_C_scalars}
\end{split}
\end{align}
Fig. \ref{fig:coloron-scalars} shows the coloron decay  branching-ratios into $G_H \phi_I$ as functions of $M_C$ and for different pseudoscalar masses and $\cot\omega$ values.
We find that $BR[C \to G_H \phi_I]$ reaches its maximum value of about 5\% for $\tan\omega\simeq 1.3$.  The decay width $ \Gamma \left[C \to G_H G_H\right]$ is negligible in the region $\cot\omega\lesssim 2$ and comparable to $\Gamma \left[C \to G_H \phi_I\right]$ for $\cot\omega> 2$.

Contrary to the case in \cite{Bai:2010dj} where the coloron decays into scalars can be relevant for $\cot\omega\gg 1$, we find that in our model these decays have branching-ratios typically below $\sim 5\%$. The difference is due to the fact that, for $\cot\omega\gg 1$, coloron decays into third-generation quarks become the dominant decay modes in our model. For $\cot\omega<1$, on the other hand, colorons decay mainly into light-quarks. Because of the small rates for the coloron decays into scalars, we conclude that the limits on $M_C$ derived in \cite{Chivukula:2013kw} neglecting the decays into scalars, as described above, are still valid in the case of light $G_H$ and $\phi_I$.

\begin{figure}
\centering
\includegraphics[width=.6\textwidth]{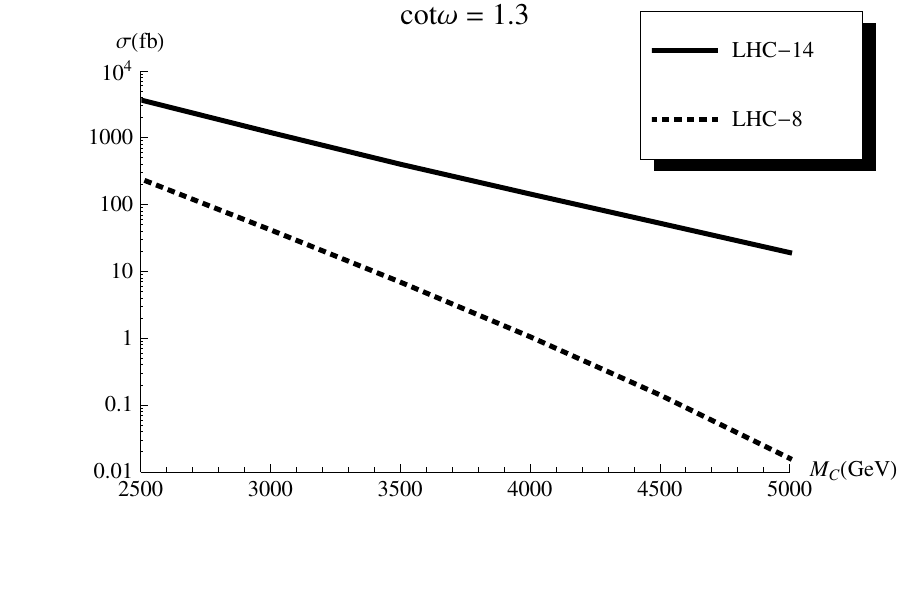}
\includegraphics[width=.7\textwidth]{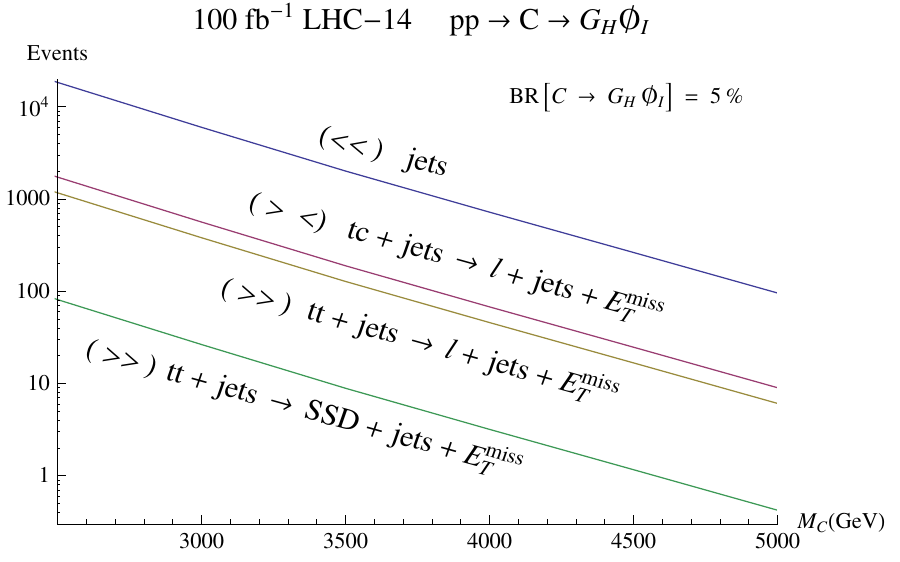}
\caption{Upper Plot: cross sections for coloron production, with $\tan\omega=1.3$, at the 8 TeV (dotted) and at the 14 TeV LHC (solid). Lower Plot: Expected number of events with 100 fb$^{-1}$ at the 14 TeV LHC for the process $pp \to C \to G_H \phi_I$ with different final states, corresponding to the most promising signatures in the different scalar mass regions: ($<<$) denotes the region where both the scalars are lighter than the top, ($>>$) the region where are both heavier, ($><$) that where one scalar is lighter than the top and the other is heavier; we assume a 5\% coloron decay branching fraction into $G_H\phi_I$.  }
\label{fig:coloron-xsec}
\end{figure}

Considering the significant rates for coloron production at the 14 TeV LHC (of the order of 10$^2$-10$^3$ fb in Fig. \ref{fig:coloron-xsec}),  a $BR[C\to G_H \phi_I]$ of the order of 5\%  offers a significant possibility for discovering the pseudoscalar $\phi_I$ in the channel $pp\to C\to G_H \phi_I$. 
These channels have a distinctive topology, characterized by the presence of three resonances, two of which are narrow, and peculiar kinematics, in which the light scalars, coming from the decay of the heavy coloron, are expected to be boosted.
The lower plot in Fig. \ref{fig:coloron-xsec} shows the expected number of events with 100 fb$^{-1}$ at the 14 TeV LHC for the process $pp \to C \to G_H \phi_I$ with different final states, corresponding to the most promising signatures in the different scalar mass regions. In the case where both $G_H$ and $\phi_I$ are lighter than the top, the scalars decay into jets \footnote{Except for the case in which the scalars are heavier than approximately 150 GeV, where they decay dominantly into an off-shell top and a charm.}, leading to dijet pair signatures. If one scalar is lighter than the top and the other is heavier, the scalar decays lead to a final state $tc+jets$, which is probably observable by considering the leptonic decay of the top.  If both the scalars are heavier than the top, the intermediate final state is $tt+jets$. In this case the most promising discovery signature is the semi-leptonic channel, where one top decays leptonically and the other hadronically. 
The cleaner single-sign dilepton channel will probably yield too few events to allow
for a discovery at 100 fb$^{-1}$, but it could be an interesting confirming signature at larger integrated luminosity.  
Detailed consideration of this signal is left to a future study.

\section{Color-Singlet Scalar Phenomenology}
\label{Sec:phiR}

Finally, we comment briefly on the phenomenology of the color-singlet scalar in this section. As we will show, the $\phi_R$ phenomenology is less interesting for direct searches at colliders than those of $G_H$ and $\phi_I$. The $\phi_R$ interactions with quarks are analogous to those of the color-octet in Fig. \ref{fig:dia-GH-coupl}, and can be obtained from those of the color-octet by substituting $T^{a} \to \mathcal{I}/\sqrt{6}$. The dominant color-singlet scalar decay modes are thus similar to those of the octect. The $\phi_R$ can be only singly produced at one loop through its interaction with a pair of octets, as generated by the $\mu$ cubic term in the scalar potential (\ref{eq:potential}). We have
\begin{equation}\label{eq:PhiR-GH}
\frac{\mu}{6}\sqrt{\frac{3}{2}} D_{ab} G^a_H G^b_H \phi_{R}~,
\end{equation}
where $D_{ab}=2 \epsilon_{ijk} \epsilon_{i^{'}j^{'}k} T^{a}_{ii^{'}} T^{b}_{jj^{'}}$ and $\sum_{a,b} D^2_{ab}=8$.
The $\phi_R$ single production can be described by the effective coupling
\begin{equation}\label{eq:single-eff-phiR}
-\frac{1}{4} C_{g \phi R} D_{ab} G^a_{\mu\nu} G^{\mu\nu b} \phi_R~,
\end{equation}
where, for $M_{\phi_R}\simeq M_{G_H}$,
\begin{equation}\label{eq:cGHgphiR}
C_{g \phi R}=\sqrt{\frac{6}{5}}\alpha_s\frac{\mu}{\pi M^2_{\phi_R}} \left(\frac{\pi^2}{9}-1\right)\ .
\end{equation}
%

We have calculated the $\phi_R$ single-production rate by implementing the above interaction (\ref{eq:single-eff-phiR}) in \uppercase{MadGraph}. We obtain cross sections of the order of 10-10$^2$ fb for $\phi_R$ masses in a range [50, 200] GeV. These are so small that it would be difficult to discover even a very light $\phi_R$ at the LHC, because of the overwhelming QCD background. For this reason, the $\phi_R$ phenomenology is less interesting for direct searches at colliders, but it could be important for indirect searches for new Physics. For example, the possible $\phi_R$ mixing with the Higgs could affect the Higgs decay rates 
\cite{Kumar:2012ww,Kribs:2012kz,Cao:2013wqa,Chang:2012ta,Dobrescu:2011aa,Manohar:2006ga,Arsham}. We leave this aspect to further investigation.

\section{Conclusions and Outlook}
\label{Sec:conclusion}

We have studied the phenomenology of color-octet and color-singlet scalars in the flavorful Top-Coloron model introduced in \cite{Chivukula:2013kw}. Our results for the octet phenomenology are summarized in Table \ref{tab:GH-pheno}, which shows, for different octet masses, the dominant decay modes, the most promising signatures for discovery at colliders and the mass regions already excluded by Tevatron and LHC. 

We find that the same-sign-dilepton final state is the `golden' channel for discovering the octet, which is mainly produced in pairs and decays dominantly, if it is heavier than the top, into $t\bar{c}$ or $\bar{t}c$. The searches for new physics at the 8 TeV LHC in this channel place a lower limit of 440 GeV on the octet mass. The case of an octet lighter than the top, where the octet completely decays into jets, has been tested by the Tevatron, which excludes the mass region from 50 to 125 GeV. Color-singlets are produced at rates much smaller than the octets and the current searches at colliders do not place direct bounds on their mass. 

Finally, we identify a promising channel for observing the color-singlet pseudoscalar at the 14 TeV LHC: the channel where the color-singlet is produced in association with a color-octet, from the decay of a coloron. Fig. \ref{fig:coloron-xsec} shows, for this process, the expected number of events with 100 fb$^{-1}$ at the 14 TeV LHC for several final states, corresponding to the most promising signatures in the different scalar mass regions.

\begin{acknowledgements}
We thank Bogdan Dobrescu for insightful comments.
R.S.C., E.H.S., and \ N.V. were supported, in part, by the U.S.\ National Science Foundation under Grant No.\ PHY-0854889.   R.S.C. and E.H.S thank the Aspen Center for Physics  and the NSF Grant No.\ 1066293  for hospitality during the completion of this work.

\end{acknowledgements}

\appendix
\section{Limits from Flavor Violation}
\label{Sec:flavor}

Here we summarize the rather weak flavor physics constraints on our model. The most dangerous flavor-violating terms involving the exchange of the color octet and singlet scalars are those inducing B-meson mixing at tree level. The scalars' exchange contributions to $\Delta F=1$ processes, as $b\to s\gamma$ or $\epsilon^{'}/\epsilon_K$, are subleading in the flavor-violating couplings or require mass insertions in the external legs and can thus be safely neglected. We do not expect large corrections even for light scalars. Indeed, the scalar exchange contributions are suppressed by the Yukawa couplings \cite{Chivukula:2013kw}. 

The $G_H$ flavor-violating interactions in Eq. (\ref{eq:GH-psi-b}), give a contribution to the four-fermion $O^{bs/bd}_2$ and $O^{bs/bd}_3$ operators \cite{Bona:2007vi}:
\begin{align}
\begin{split}
& O^{bs}_2  = \bar{b}^{i}_R s^{i}_L\bar{b}^{j}_R s^{j}_L \qquad O^{bd}_2  = \bar{b}^{i}_R d^{i}_L\bar{b}^{j}_R d^{j}_L \\
& O^{bs}_3  = \bar{b}^{i}_R s^{j}_L\bar{b}^{j}_R s^{i}_L \qquad O^{bd}_2  = \bar{b}^{i}_R d^{j}_L\bar{b}^{j}_R d^{i}_L
\end{split}
\end{align}
where $i,j$ are color indexes. 
The absolute value of the 
$G_H$ contribution to $O^{bs}_2$ reads:

\begin{equation}
\left|C^{B_s}_2 \right|=\frac{1}{6}\frac{1}{M^2_{G_H}}\frac{M^2}{u^2}\left|\alpha_2 \lambda^{'}_b\right|^2 \leq \frac{1}{6}\frac{1}{M^2_{G_H}}\frac{m^2_b}{u^2}\left(0.0085\right)^2
\end{equation}
where we have used the constraint $|\alpha_2 \lambda^{'}_b | / \beta_b \leq 0.0085$, coming from the limit on the corrections to $b \to s \gamma$, and the estimate $\beta_b \simeq m_b/M$.
The study in \cite{Bona-vietnam} fixes the bound, $\left|C^{B_s}_2 \right| < 1.1 \cdot 10^{-11}$ GeV$^{-2}$, from which, taking $|\alpha_2 \lambda^{'}_b | / \beta_b = 0.0085$, we extract the limit:
\begin{equation}\label{eq:FV-limit}
M_{G_H} > 1.1 \, m_b \, \left[\frac{1\, \text{TeV}}{u}\right] \ .
\end{equation}
The coefficient $C^{B_d}_2$ can be obtained from $C^{B_s}_2$ by making the substitution $\alpha_2 \to \alpha_1$. The resulting bound $\left|C^{B_d}_2 \right| < 4.4 \cdot 10^{-13}$ leads to a milder constraint, $M_{G_H}> 0.44 \, m_b$ [TeV/$u$].
The Wilson coefficients $|C^{B_{s/d}}_3|$ are obtained through the relation $|C^{B_{s/d}}_3| = 3\, |C^{B_{s/d}}_2|$. The corresponding limits $|C^{B_{s}}_3|< 3.9 \cdot 10^{-9}$ and $|C^{B_{d}}_3|< 1.5 \cdot 10^{-12}$ lead, again, to constraints milder than that in (\ref{eq:FV-limit}); we obtain $M_{G_H} > 0.93\, (0.41) \, m_b \, [\text{TeV}/u]$ from the bound on $C^{B_{s}}_3 \ (C^{B_{d}}_3)$.

The color-singlet scalars $\phi_I$ and $\phi_R$ give contributions only to $O^{bs/bd}_2$. In fact, the absolute value of the Wilson coefficients of the $\phi_I$ and $\phi_R$ contributions to $O^{bs}_2$ are the same as those from $G_H$;  there is no color factor  of $1/6$ but there is a compensating new factor $(1/\sqrt{6})^2$ within the flavor-violating couplings of the color-singlet scalars.
We thus arrive at the bound:

\begin{equation}\label{eq:FV-limit}
M_{\phi_I} > 1.1 \, m_b \, \left[\frac{1\, \text{TeV}}{u}\right] \ .
\end{equation}


\end{document}